\newif\ifarxiv
\arxivtrue % on arxiv 
\ifarxiv
	\documentclass[10pt,twocolumn,journal]{IEEEtran}	 
\else	
    \documentclass{ieeeaccess}	
\fi

\newif\ifhighlight
\highlightfalse % do not highlight revised text

\ifhighlight
\newcommand{\revisedtext}[1]{\hl{#1}}
\else
\newcommand{\revisedtext}[1]{#1}
\fi

\usepackage{hyperref}
\usepackage{graphicx} % Include figure files
\usepackage{dcolumn} % Align table columns on decimal point
\usepackage{bm} % bold math
\usepackage{subcaption} 
\usepackage{color,soul}
\usepackage{multirow}
\usepackage[version=3]{mhchem}
\usepackage{url}
\usepackage{amssymb}

\ifarxiv
\else % Fix problem with IEEE Access class with mhchem
	\usepackage{tikz} 
	\NewSpotColorSpace{PANTONE}
	\AddSpotColor{PANTONE} {PANTONE3015C} {PANTONE\SpotSpace 3015\SpotSpace C} {1 0.3 0 0.2}
	\SetPageColorSpace{PANTONE}%
\fi

\newtheorem{remark}{Remark}

\begin{document}

\title{Using biochemical circuits to approximately compute log-likelihood ratio for detecting persistent signals}

\ifarxiv
\author{Chun Tung Chou \\
School of Computer Science and Engineering, University of New South Wales, Sydney, New South Wales 2052, Australia. \\
E-mail: c.t.chou@unsw.edu.au}

\else
%\author{Chun Tung Chou,~\IEEEmembership{Member,~IEEE,}
%\IEEEcompsocitemizethanks{\IEEEcompsocthanksitem C.T. Chou is with the School of Computer Science and Engineering, University of New South Wales, Sydney, New South Wales 2052, Australia. E-mail: c.t.chou@unsw.edu.au \protect
%}}
\history{Date of publication xxxx 00, 0000, date of current version xxxx 00, 0000.}
\doi{10.1109/ACCESS.2017.DOI}

\author{\uppercase{Chun Tung Chou}\authorrefmark{1}, \IEEEmembership{Member, IEEE}}
\address[1]{School of Computer Science and Engineering, University of New South Wales, Sydney, New South Wales 2052, Australia (e-mail: c.t.chou@unsw.edu.au)}
% \tfootnote{}

%\markboth
%{Author \headeretal: }
%{Author \headeretal: }

\corresp{Corresponding author: Chun Tung Chou (e-mail: c.t.chou@unsw.edu.au).}

\markboth{Submitted to IEEE Access \today}{}
\fi
% \orcid{0000-0003-4512-7155}

\ifarxiv  % The placement of \maketitle depends on the set-up 
\maketitle
\fi

\begin{abstract}
Given that biochemical circuits can process information by using analog computation, a question is: What can biochemical circuits compute? This paper considers the problem of using biochemical circuits to distinguish persistent signals from transient ones. We define a statistical detection problem over a reaction pathway consisting of three species: an inducer, a transcription factor (TF) and a gene promoter, where the inducer can activate the TF and an active TF can bind to the gene promoter. We model the pathway using the chemical master equation so the counts of bound promoters over time is a stochastic signal. We consider the problem of using the continuous-time stochastic signal of the counts of bound promoters to infer whether the inducer signal is persistent or not. We use statistical detection theory to derive the solution to this detection problem, which is to compute the log-likelihood ratio of observing a persistent signal to a transient one. We then show, using time-scale separation and other assumptions, that this log-likelihood ratio can be approximately computed by using the continuous-time signals of the number of active TF molecules and the number of bound promoters when the input is persistent. Finally, we show that the coherent feedforward gene circuits can be used to approximately compute this log-likelihood ratio when the inducer signal is persistent.  
\end{abstract}

\ifarxiv
\noindent{\bf Keywords:}
\else
\begin{keywords}
\fi

Statistical signal processing, signal detection, molecular computing, analog computation, biochemical circuits.

\ifarxiv
\else
\end{keywords} 
\maketitle
\fi

\section{Introduction}
The fact that living organisms use biomolecular circuits for information processing has provided much inspiration for engineers to design new biomolecular circuits. Firstly, engineers have been inspired to use engineering theory to design synthetic circuits. E.g., control theory has been used to design circuits for homeostatic control \cite{Teo:2019bk}, concentration regulation \cite{Pasotti:2019fi} and to counter the effect of loading \cite{McBride:2019is}; signal processing theory has been used to design molecular circuits that can be used for communications \cite{Kuscu:2019ey,Chou:2019gf,Awan:2018ig} and filtering \cite{Zechner:2016fb}. Secondly, one can view the information processing carried out by biomolecular circuits as analog computation \cite{Sarpeshkar:2014dg,Teo:2015kva}. This view point has helped engineers to design molecular circuits that can perform logarithmic sensing \cite{Daniel:2013ke}, parity check \cite{Marcone:2018kp} and integral control \cite{Briat:2016ds}. This paper considers the problem of persistence detection, i.e. the problem of deciding whether a particular chemical species has been present in sufficient quantity for a long enough time. In the natural world, the bacteria {\sl Escherichia coli} ({\sl E.~coli}) perform persistence detection \cite{Mangan:2003ia}. In the synthetic world, cell-based therapy \cite{Mimee:2015jo} can make use of persistence detection on biomarkers. In this paper, we will solve the persistence detection problem by using statistical detection theory \cite{Kay_v2} and show that a gene circuit (which is specific type of biochemical circuit) can be used to approximately compute the solution of this detection problem. 

A gene circuit that can perform persistence detection in {\sl E.~coli} is the {\sl Coherent Type-1 Feedforward Loop with an AND logic at the output} \cite{ShenOrr:2002jo} or the C1-FFL for short. The C1-FFL is a network motif and is a frequently found circuit in both {\sl E.~coli} and {\sl Saccharomyces cerevisiae} (yeast) \cite{Milo:2002cg,ShenOrr:2002jo}. This means that the C1-FFL carries out important functions in cells. The authors in \cite{ShenOrr:2002jo} showed that the C1-FFL can act as a persistence detector. They did this by modelling the gene expression in the C1-FFL by using ordinary differential equations (ODEs) and show that a persistent (resp. transient) input to the C1-FFL will result in a high (zero) output. 

The paper \cite{ShenOrr:2002jo} took a deterministic approach to understand persistence detection. Given that the biochemical environment is stochastic, it is therefore necessary to understand how cells can infer information on the environment from a stochastic point of view \cite{Libby:2007fw}. This paper considers a reaction pathway consisting of three chemical species: an inducer, a transcription factor (TF) and a gene promoter. In this reaction pathway, the inducer can activate the TF and the activated TF can bind with the gene promoter. In order to model a stochastic biochemical environment, we model the reaction pathway using the chemical master equation \cite{Gardiner}. We consider a detection problem whose aim is to infer whether the inducer signal is persistent or not by using the signal of the number of bound promoters over time.  According to detection theory, the solution to this detection problem is to compute a log-likelihood ratio and we derive an ODE which describes the evolution of this log-likelihood ratio over time. In order to connect this ODE to the C1-FFL, we use time-scale separation and other assumptions to derive an intermediate approximation which is an ODE that can approximately compute the log-likelihood ratio for persistent signals. We then show that this intermediate approximation can be realised by using a C1-FFL. \revisedtext{The key contribution of this paper is to show that it is possible to find the parameters of a C1-FFL so that its output is approximately equal to the log-likelihood ratio of statistical detection problem when the inducer signal is persistent. More specifically, the aim of this statistical detection problem is to detect whether the inducer signal is persistent in an inducer-TF-gene pathway.} In addition, the methodology in this paper can be useful for designing synthetic molecular circuits for performing other signal processing tasks. 

\revisedtext{
This paper makes advances compared to our previous work {\cite{Chou:2018jh}}. In comparison to  {\cite{Chou:2018jh}}, this paper makes two different assumptions: (i) This paper considers an inducer-TF-gene pathway but {\cite{Chou:2018jh}} considered only an inducer-TF pathway; (ii) This paper assumes that the inducer signal is stochastic while {\cite{Chou:2018jh}} assumed that the inducer signal is deterministic. These two different assumptions mean that new methodologies are needed to show that the C1-FFL can be used to approximately compute the log-likelihood ratio of a persistence detection problem. First, we need to show how the log-likelihood ratio can be computed exactly (Sec.~{\ref{sec:dp:sol}}) in an inducer-TF-gene pathway. Second, we need to derive a new method to approximately compute the log-likelihood ratio (Sec.~{\ref{sec:approx_llr}}). In particular, this paper needs to approximate the solution to a Bayesian filtering problem {\cite{Bronstein:2018eh}} but this is not required in {\cite{Chou:2018jh}} as it considered a deterministic inducer signal. Third, we need to derive a method to show how the parameters of the approximate log-likelihood ratio computation can be mapped to the C1-FFL parameters (Sec.~{\ref{sec:c1ffl}}). These three aspects are the new elements of this paper in comparison to  {\cite{Chou:2018jh}}. Furthermore, in comparison with our earlier conference paper {\cite{Chou:2019ur}}, this paper explains why a C1-FFL can be used to approximately compute the approximate log-likelihood ratio  (Sec.~{\ref{sec:c1ffl}}) and provide full derivation on the computation of exact and approximate log-likelihood ratios (Appendices {\ref{app:sol:dp}} and {\ref{app:ia}}). 
}

The rest of this paper is organised as follows. Sec.~\ref{sec:bg} presents background information on the C1-FFL. We then define the detection problem and present its solution in Sec.~\ref{sec:results}. After that in Sec.~\ref{sec:approx_llr}, we present a method to approximately compute the log-likelihood ratio and use this approximation in Sec.~\ref{sec:c1ffl} to show that the C1-FFL can be used to approximately compute the the log-likelihood ratio when it is positive. Finally, Sec.~\ref{sec:final} presents a discussion and concludes the paper.

\section{Background on C1-FFL}
\label{sec:bg}

\begin{figure}[t]
        \centering
        \includegraphics[page=3,scale=0.35,trim={1cm 2cm 21cm 2cm},clip]{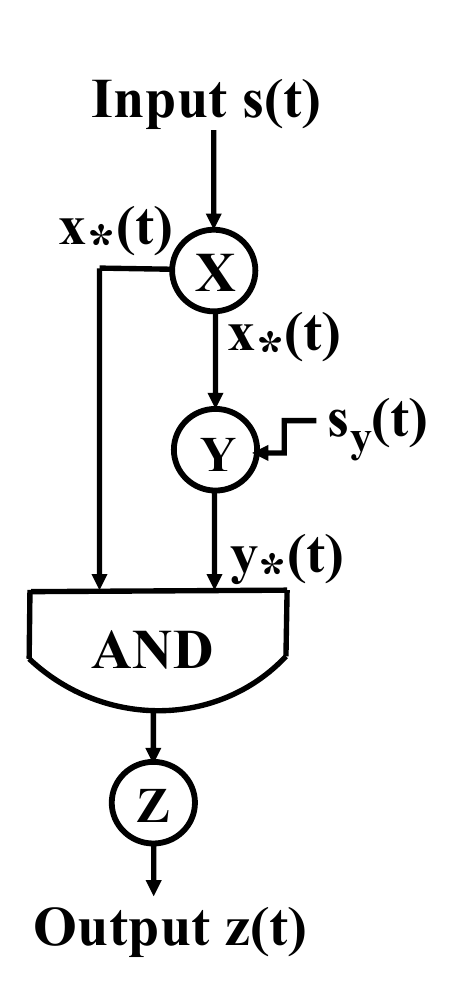} 
        \caption{Network representation of C1-FFL.}
        \label{fig:c1ffl_a}
\end{figure}

\begin{figure}[t]
        \centering
        \includegraphics[page=1,scale=0.30,trim={1cm 10cm 2cm 2cm},clip]{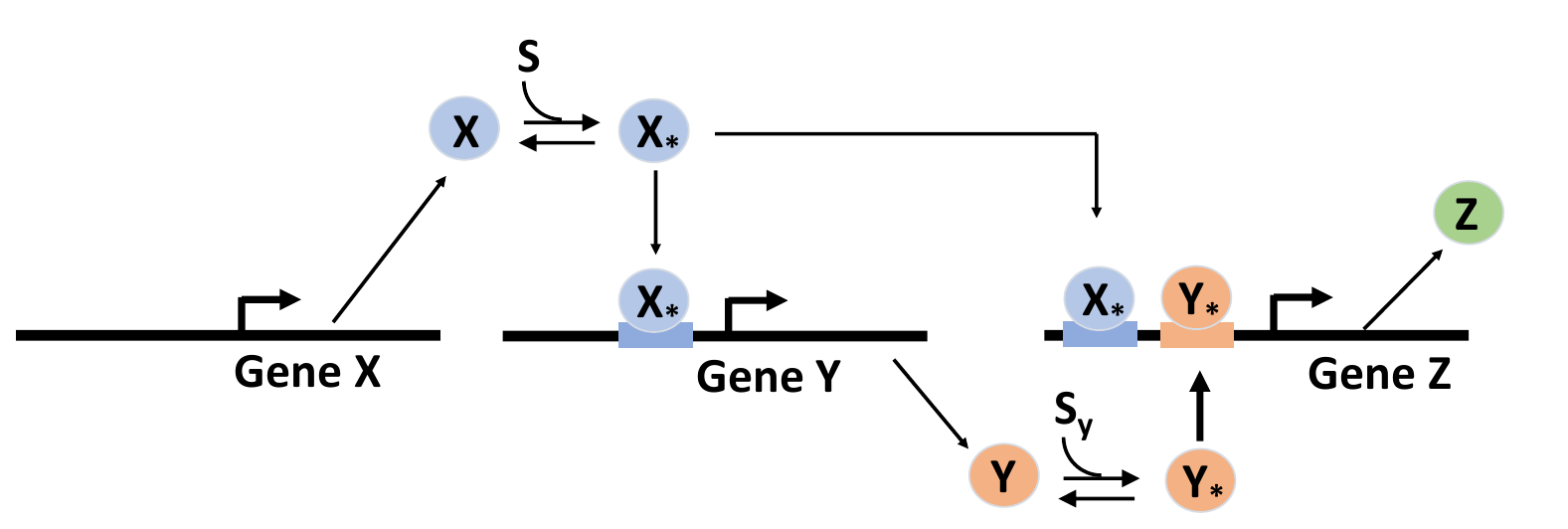}
        \caption{Representation of C1-FFL using inducers, transcription factors and genes.}
        \label{fig:c1ffl_b}
\end{figure}

The C1-FFL can be depicted as a network where each link is associated with a signal and each node transforms the input signal(s) into an output signal. Fig.~\ref{fig:c1ffl_a} shows the network of the C1-FFL. The input signal is $s(t)$ and output signal is $z(t)$. Both $x_*(t)$ and $y_*(t)$ are intermediate signals, and $s_y(t)$ is an external signal. 

The C1-FFL in Fig.~\ref{fig:c1ffl_a} is an abstraction of the molecular interactions which are depicted in Fig.~\ref{fig:c1ffl_b}. In the figure, both S and \cee{S_y} are inducers. Both X and Y are TFs, which are expressed by their corresponding gene. The inducer S (resp. \cee{S_y}) turns the inactive form  X (Y) into its active form \cee{X_*} (\cee{Y_*}). The activation of gene Z requires the binding of both \cee{X_*} and \cee{Y_*} to the promoter of Z, i.e. the AND gate in Fig.~\ref{fig:c1ffl_a}.

Note that there is a one-to-one correspondence between the chemical species in Fig.~\ref{fig:c1ffl_b} with their corresponding time signals in Fig.~\ref{fig:c1ffl_a}, e.g. $x_*(t)$ is the concentration of \cee{X_*} at time $t$ and so on. In this paper, we will assume that the inducer \cee{S_y} is always present and its concentration is always above the threshold needed to activate Y. Furthermore, we assume the activation of Y by \cee{S_y} is fast, this allows us to write $y_*(t) = y(t)$ and we will use $y(t)$ for $y_*(t)$ from now on. By using Hill function to model the gene expression,  \cite{Mangan:2003ia} presents an ODE model for the C1-FFL, as follows:
\begin{subequations}
\label{eq:ffl_all}
\begin{align}
\frac{dx_*(t)}{dt} =& k_+ (M - x_*(t)) s(t) - k_- x_*(t) \label{eq:ffl1} \\
\frac{dy(t)}{dt} =&  \underbrace{\frac{h_{xy} x_*(t)^{n_{xy}}}{K_{xy}^{n_{xy}} + x_*(t)^{n_{xy}}}}_{H_{xy}(x_*(t))} - d_y y(t) \label{eq:ffl2} \\
\frac{dz(t)}{dt} =&  \underbrace{\frac{h_{xz} x_*(t)^{n_{xz}}}{K_{xz}^{n_{xz}} + x_*(t)^{n_{xz}}}}_{H_{xz}(x_*(t))}  \times \underbrace{\frac{h_{yz} y(t)^{n_{yz}}}{K_{yz}^{n_{yz}} + y(t)^{n_{yz}}}}_{H_{yz}(y(t))}  - d_z z(t)  \label{eq:ffl3} 
\end{align}
\end{subequations}
where $k_+$, $k_-$, $d_y$ and $d_z$ are reaction rate constants; $h_{xy}$, $n_{xy}$, $K_{xy}$, $h_{xz}$, $n_{xz}$, $K_{xz}$, $h_{yz}$, $n_{yz}$ and $K_{yz}$ are coefficients for Hill functions $H_{xy}(x_*(t))$, $H_{xz}(x_*(t))$ and $H_{yz}(y(t))$. Lastly, $x(t)+x_*(t)$ is the constant $M$. The multiplication of $H_{xz}$ and $H_{yz}$ on the right-hand side (RHS) of \eqref{eq:ffl3} implements the AND gate in Fig.~\ref{fig:c1ffl_a}. With suitably chosen parameter values, the C1-FFL in \eqref{eq:ffl_all} acts as a persistence detector in the sense that if the input signal $s(t)$ is a persistent (resp. transient), then the output $z(t)$ has a high (low) value.

\section{Statistical detection on a reaction pathway} 
\label{sec:results} 
Our aim is to consider a statistical detection problem to determine whether the input signal is persistent or not. However, in this section, we will consider a more general detection problem because it can readily be solved and we will specialise it to persistence detection in Sec.~\ref{sec:approx_llr}. This section is divided into two parts. We define the detection problem in Sec.~\ref{sec:dp:def} and present its solution in Sec.~\ref{sec:dp:sol}. 

{\bf Convention:} In this paper, we use upper case letters to denote a chemical species, e.g. S, \cee{X_*} etc. For each chemical species, there are two corresponding continuous-time signals based on its concentration and molecular counts. E.g. for the chemical species \cee{X_*}, we denote its concentration over time as $x_*(t)$ (note: lower case $x$) and its molecular counts over time is $X_*(t)$ (note: upper case $X$).  

\subsection{Detection problem}
\label{sec:dp:def} 

\begin{figure}[t]
    \centering
        \includegraphics[page=2,scale=0.35,trim={1.5cm 8cm 3.5cm 2cm},clip]{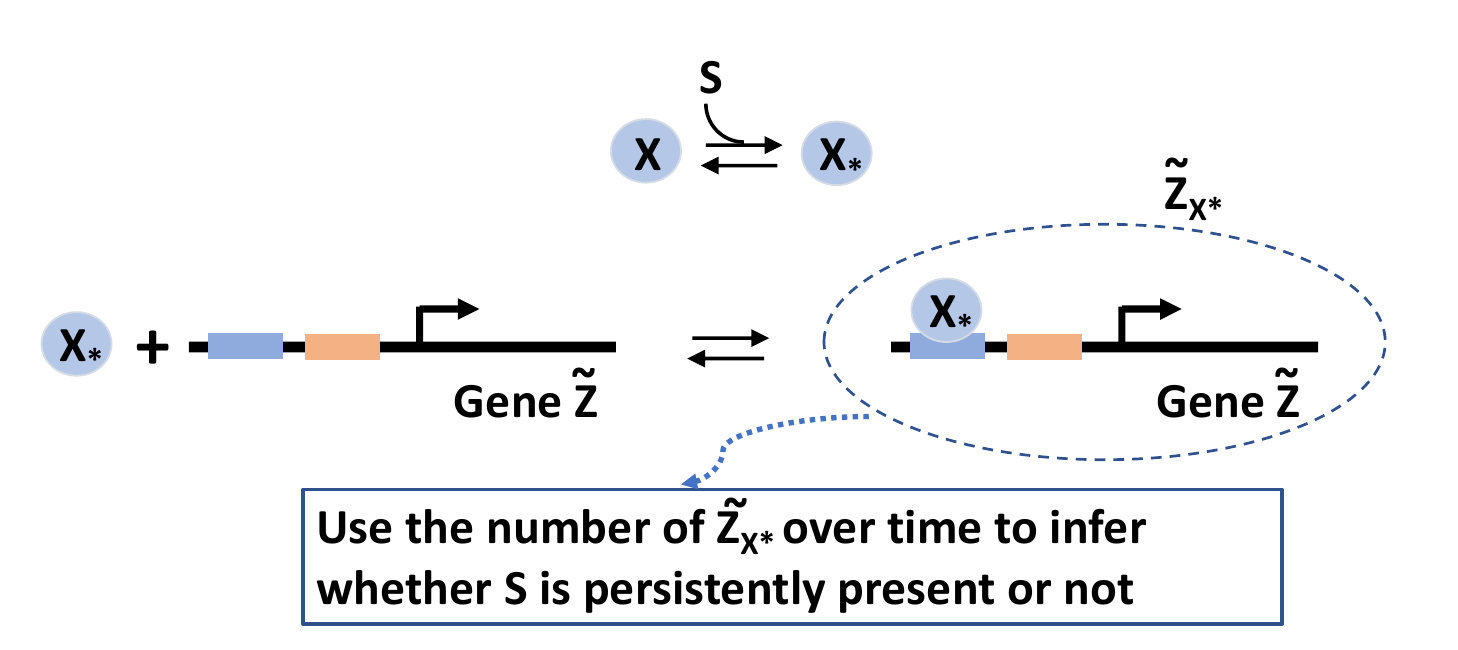}   
        \caption{The reaction pathway for the detection problem.}
        \label{fig:dp:pathway}
\end{figure}

In order that we can connect the detection problem to the C1-FFL later on, we will define the detection problem using a reaction pathway which is a {\sl subset} of the C1-FFL species and reactions in Fig.~\ref{fig:c1ffl_b}. We have depicted the reaction pathway used in the detection problem in Fig.~\ref{fig:dp:pathway}. The reaction pathway consists of five chemical species: S, inactive \cee{X} and its corresponding active form \cee{X_*}, as well as inactive $\widetilde{\rm Z}$ and the complex $\widetilde{\rm Z}_{X_*}$ which is formed by the binding of \cee{X_*} to $\widetilde{\rm Z}$. These five species take part in the following four chemical reactions:
\begin{subequations}
\label{cr:z_all} 
\begin{align}
\cee{
S + X &  ->[k_+] S + X_* \label{cr:on1}  \\
X_* &  ->[k_-] X \label{cr:off1} \\
X_* +  \widetilde{\rm Z} &  ->[g_+]   \widetilde{\rm Z}_{X_*} \label{cr:on2}  \\
\widetilde{\rm Z}_{X_*} &  ->[g_-]  X_* + \widetilde{\rm Z}   \label{cr:off2}}
\end{align}
\end{subequations}
where $k_+$, $k_-$, $g_+$ and $g_-$ are reaction propensity constants. For the time being, we will make the simplifying assumption that the volume scaling needed to convert between propensity and reaction rate constants is 1. This simplification allows us to equate propensity constants with reaction rate constants. We will explain how non-unit volume can be dealt with in Remark \ref{re:v}. With this assumption, note that $k_+$ and $k_-$ in \eqref{cr:on1} and \eqref{cr:off1} are equal to those in \eqref{eq:ffl1}. 

In terms of molecular biology, \cee{S} is an inducer and \cee{X} is a TF. In Reaction \eqref{cr:on1}, the species \cee{S} activates \cee{X} to produce \cee{X_*}. Reaction \eqref{cr:off1} is a deactivation reaction. The reactions \eqref{cr:on1} and \eqref{cr:off1} are depicted in both Figs.~\ref{fig:c1ffl_b} and \ref{fig:dp:pathway}. 

The species $\widetilde{\rm Z}$ is a gene. In fact, $\widetilde{\rm Z}$ in Fig.~\ref{fig:dp:pathway} is the same as \cee{Z} in Fig.~\ref{fig:c1ffl_b}. Note that Fig.~\ref{fig:c1ffl_b} follows the standard convention in molecular biology where a gene and the protein that it expresses are given the same symbol \cee{Z}. However, in this paper, we need different symbols for the gene and the protein that the gene expresses so that we can clearly distinguish their corresponding time signals. Therefore, we have chosen to use $\widetilde{\rm Z}$ to denote the gene and use \cee{Z} to denote the protein expressed by $\widetilde{\rm Z}$. In Reaction \eqref{cr:on2}, an active \cee{X_*} binds with the promoter of $\widetilde{\rm Z}$ to produce the complex $\widetilde{\rm Z}_{X_*}$. Lastly, Reaction \eqref{cr:off2} is an unbinding reaction. 

% Note that we have intentionally chosen the reaction between S and X as an approximate enzymatic reaction rather than a binding reaction to simplify the detection problem solution and the discussion in this paper. We note that this approximation will not affect the results in this paper. %  if the amount of S is much larger than that of X, which is usually the case in living cells. 

Let $S(t)$, $X(t)$, $X_*(t)$, $\widetilde{Z}(t)$ and $\widetilde{Z}_{X_*}(t)$ denote, respectively, the {\sl number} of \cee{S}, \cee{X},  \cee{X_*}, $\widetilde{\cee Z}$ and $\widetilde{\rm Z}_{X_*}$ molecules at time $t$. Note these signals are piecewise constant because they are molecular counts. We assume that $X(t) + X_*(t)$ (resp. $\widetilde{Z}(t) + \widetilde{Z}_{X_*}(t)$) is a constant for all $t$ and we denote this constant by $M$ ($N$). 

We will refer to $S(t)$ as the input signal. The goal of the detection is to use the signal $\widetilde{Z}_{X_*}(t)$ to determine whether the input is persistent or not. We assume that the signal $S(t)$ is generated by some species and chemical reactions upstream of \eqref{cr:z_all}. We will model these upstream reactions and \eqref{cr:z_all} using the chemical master equation \cite{Gardiner} which is a specific type of continuous-time Markov chain. We further assume that the upstream species do not react with \cee{X}, \cee{X_*}, $\widetilde{\cee Z}$ and $\widetilde{\rm Z}_{X_*}$. Intuitively, this means we can predict the behaviour of \cee{X}, \cee{X_*}, $\widetilde{\cee Z}$ and $\widetilde{\rm Z}_{X_*}$ from that of \cee{S}. 

%We assume that the chemical species \cee{S}, \cee{X}, \cee{X_*}, $\widetilde{\cee Z}$ and $\widetilde{\rm Z}_{X_*}$, as well as the reactions \eqref{cr:z_all}, are part of a ``chemical subuniverse". This subuniverse will also contain the chemical species and reactions that produce \cee{S}. We will model this subuniverse by using the chemical master equation \cite{Gardiner} which is a specific type of continuous-time Markov chain. 

We have now defined the reaction pathway and its model. The next task is to specify the measured data and the hypotheses for the detection problem. The measured datum at time $t$ is $\widetilde{Z}_{X_*}(t)$. However, in the formulation of the detection problem, we will assume that at time $t$, the data available to the detection problem are $\widetilde{Z}_{X_*}(\tau)$ for all $\tau \in [0,t]$; in other words, the data are continuous in time and are the history of the counts of $\widetilde{\rm Z}_{X_*}$ up to time $t$ inclusively. We will use ${\widetilde{\cal Z}}_{X_*}(t)$ to denote the continuous-time history of $\widetilde{Z}_{X_*}(t)$ up to time $t$ inclusively. 

We now specify the hypotheses ${\cal H}_i$ $(i = 0,1)$ for the detection problem. Later on, we will identify ${\cal H}_0$ and ${\cal H}_1$ with, respectively, transient and persistent signals. However, at this stage, we want to solve the detection problem in a general way. We assume that ${\cal H}_0$ and ${\cal H}_1$ are two distinct subsets of the set of all possible $S(t)$. Intuitively, the aim of the detection problem is to decide which signal class ${\cal H}_0$ or ${\cal H}_1$ is more likely to have produced the observed history. % Since the hypotheses are defined in terms of $S(t)$, we will refer to it as the input signal. 

We remark that in the definition of the detection problem, the input signal $S(t)$ is not directly observable. Since \cee{S} reacts with the molecules in the reaction pathway in Fig.~\ref{fig:dp:pathway}, the downstream signal $\widetilde{Z}_{X_*}(t)$ contains information on $S(t)$. The aim of the detection problem is to infer the information on $S(t)$ from this downstream signal. Given that we model the chemical system with the chemical master equation, both signals $S(t)$ and $\widetilde{Z}_{X_*}(t)$ are noisy.

\subsection{Solution to the detection problem} 
\label{sec:dp:sol}

\begin{figure}[t] % {0.35\textwidth}
        \centering
        \includegraphics[scale=0.29]{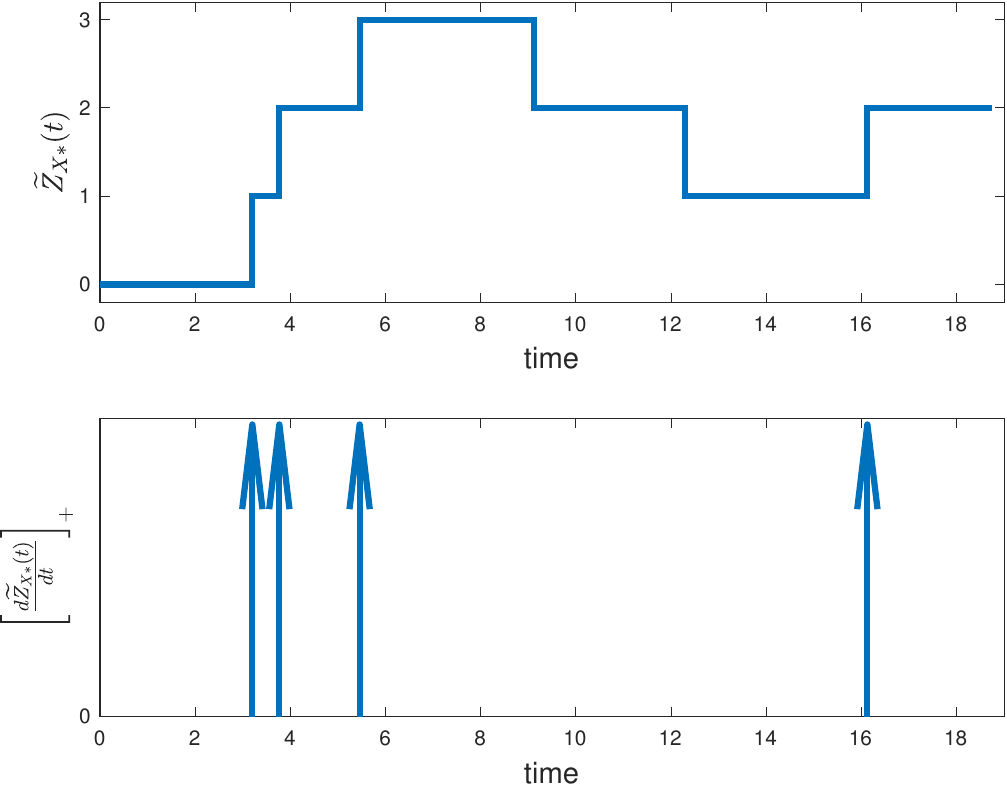}
        \caption{Illustrating $\widetilde{Z}_{X_*}(t)$ and $\left[ \frac{d\widetilde{Z}_{X_*}(t)}{dt} \right]_+$. }
        \label{fig:xstar_deri}
\end{figure}

The aim of the detection problem is to decide which hypothesis ${\cal H}_i$ $(i = 0,1)$ is more likely to have generated the observed history ${\widetilde{\cal Z}}_{X_*}(t)$. 
\revisedtext{
We assume that the two hypotheses are equally likely, therefore the log-ratio of posteriori probabilities  
$\frac{ {\rm P}[ {\cal H}_1 |   {\widetilde{\cal Z}}_{X_*}(t) ] }{  {\rm P}[ {\cal H}_0 |  {\widetilde{\cal Z}}_{X_*}(t)]   }$ 
is equal to
} the log-likelihood ratio $L(t)$:
\begin{align}
L(t) = \log\left( \frac{{\rm P}[{\widetilde{\cal Z}}_{X_*}(t) | {\cal H}_1]}{{\rm P}[{\widetilde{\cal Z}}_{X_*}(t) | {\cal H}_0]} \right)
\label{eq:LLR_t}
\end{align} 
where ${\rm P}[{\widetilde{\cal Z}}_{X_*}(t) | {\cal H}_i]$ is the conditional probability of observing the history ${\widetilde{\cal Z}}_{X_*}(t)$ given hypothesis ${\cal H}_i$.

In Appendix \ref{app:sol:dp}, we show that the time evolution of $L(t)$ is given by the following ODE:
\begin{align}
 \frac{dL(t)}{dt} =& \left[ \frac{d\widetilde{Z}_{X_*}(t)}{dt} \right]_+ \log\left(\frac{J_1(t_-)}{J_0(t_-)} \right)  -  \nonumber  \\ 
 & g_+ (N - \widetilde{Z}_{X_*}(t)) (J_1(t)-J_0(t))        \label{eq:L} \\
J_i(t) =& \; {\rm E}[X_*(t) | {\widetilde{\cal Z}}_{X_*}(t), {\cal H}_i] \label{eq:bfiltering}
\end{align}
where $[w]_+ = \max(w,0)$, ${\rm E}[\;]$ denotes the expectation and ${\rm E}[X_*(t) | {\widetilde{\cal Z}}_{X_*}(t), {\cal H}_i]$ is the conditional expectation of $X_*(t)$ given the history and ${\cal H}_i$. Note that in deriving \eqref{eq:L}, we assume that the hypotheses have been properly chosen so that \revisedtext{$J_i(t) > 0$ (for $t > 0$ and $i = 0, 1$) which in turn implies that} $\log\left(\frac{J_1(t)}{J_0(t)} \right)$ is well defined. 

Since $\widetilde{Z}_{X_*}(t)$ is a piecewise constant function counting the number of $\widetilde{\rm Z}_{X_*}$ molecules, its derivative is a sequence of Dirac deltas at the time instants that $\widetilde{\rm Z}_{X_*}$ forms or unbinds. Note that the Dirac deltas corresponding to the formation of $\widetilde{\rm Z}_{X_*}$ carries a positive sign and the $[  \; ]_+$ operator keeps only these. Fig.~\ref{fig:xstar_deri} shows an example $\widetilde{Z}_{X_*}(t)$ and its corresponding $\left[ \frac{d\widetilde{Z}_{X_*}(t)}{dt} \right]_+$. 

At the time instant $t$ that $\widetilde{\rm Z}_{X_*}$ is formed or unbind, the expectation $J_i(t)$ also has a jump in value at time $t$. The term $J_i(t_-)$ in \eqref{eq:L} refers to the value of $J_i(t)$ just before the jump. Lastly, we assume that the two hypotheses are {\sl a priori} equally likely, so $L(0) = 0$. 

We next present a numerical example to illustrate the properties of \eqref{eq:L} and to explain what information is important for persistence detection. This example will also provide some intuition on how we will approximately compute the log-likelihood in Sec.~\ref{sec:approx_llr}. 

\subsubsection{Numerical example} 
\label{sec:exact:example}

\begin{figure*}[t]
    \begin{subfigure}[t]{0.35\textwidth}
        \centering
        \includegraphics[scale=0.29]{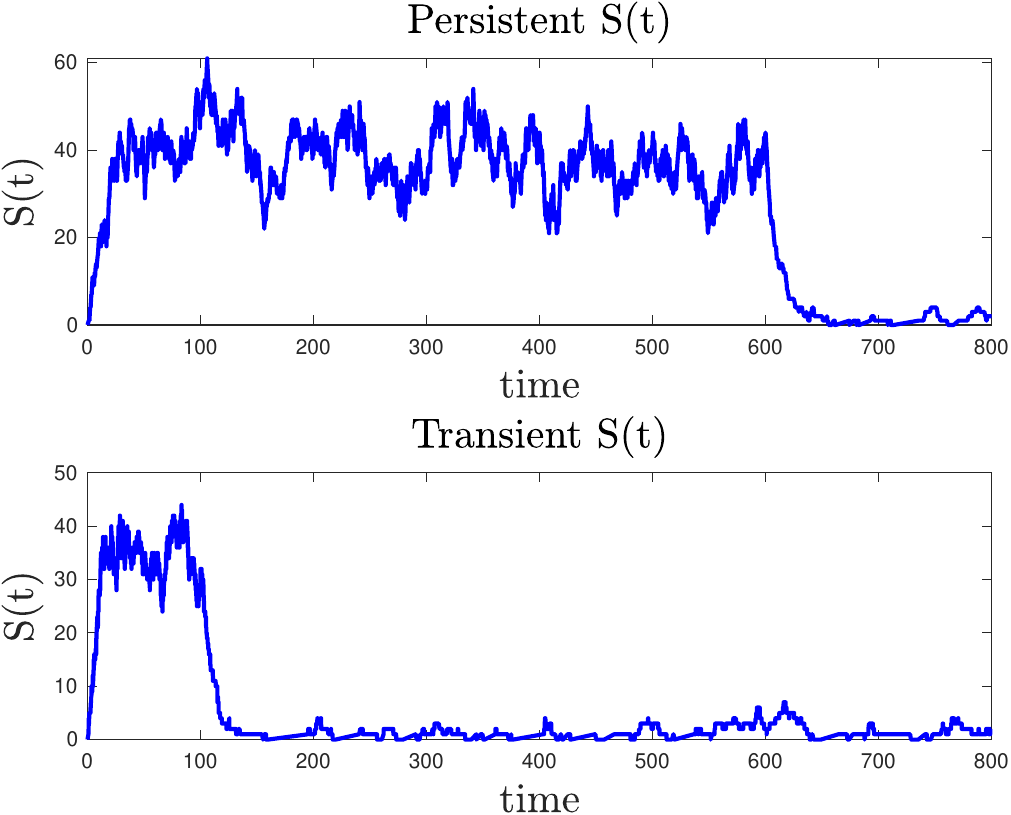}
        \caption{}
        \label{fig:LLRdemo_St}
    \end{subfigure}      
    \begin{subfigure}[t]{0.35\textwidth}
        \centering
        \includegraphics[scale=0.29]{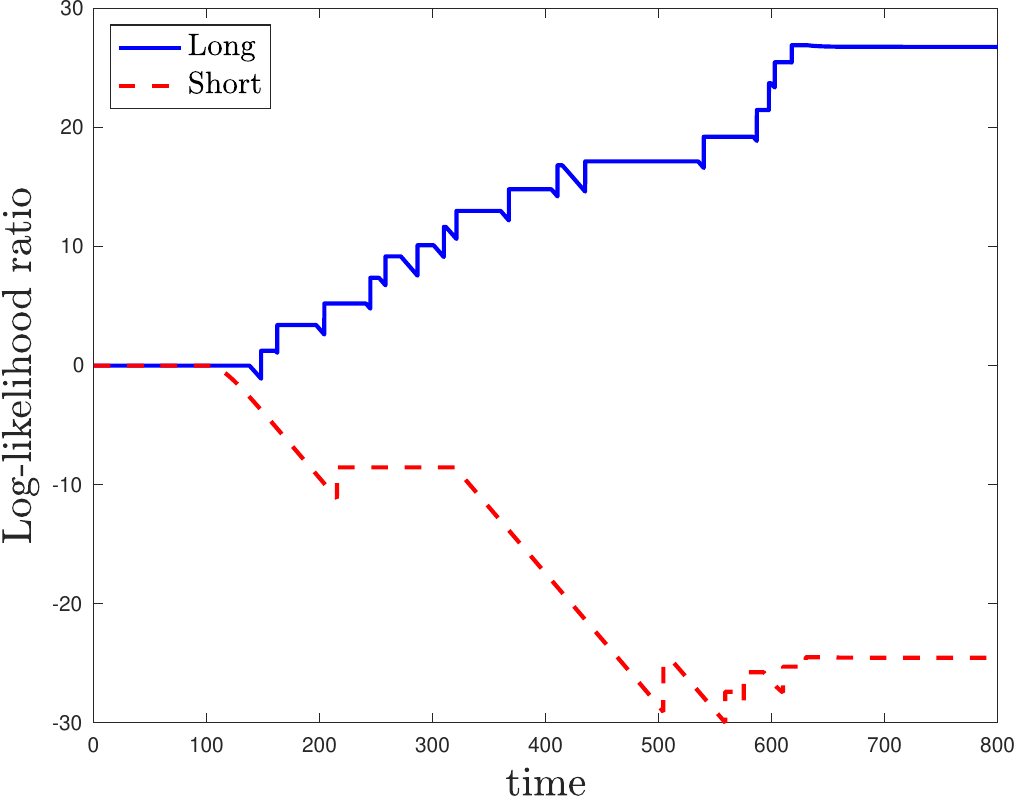}
        \caption{}
        \label{fig:LLRdemo}
    \end{subfigure}    
    \begin{subfigure}[t]{0.35\textwidth}
    \centering
    \includegraphics[scale=0.29]{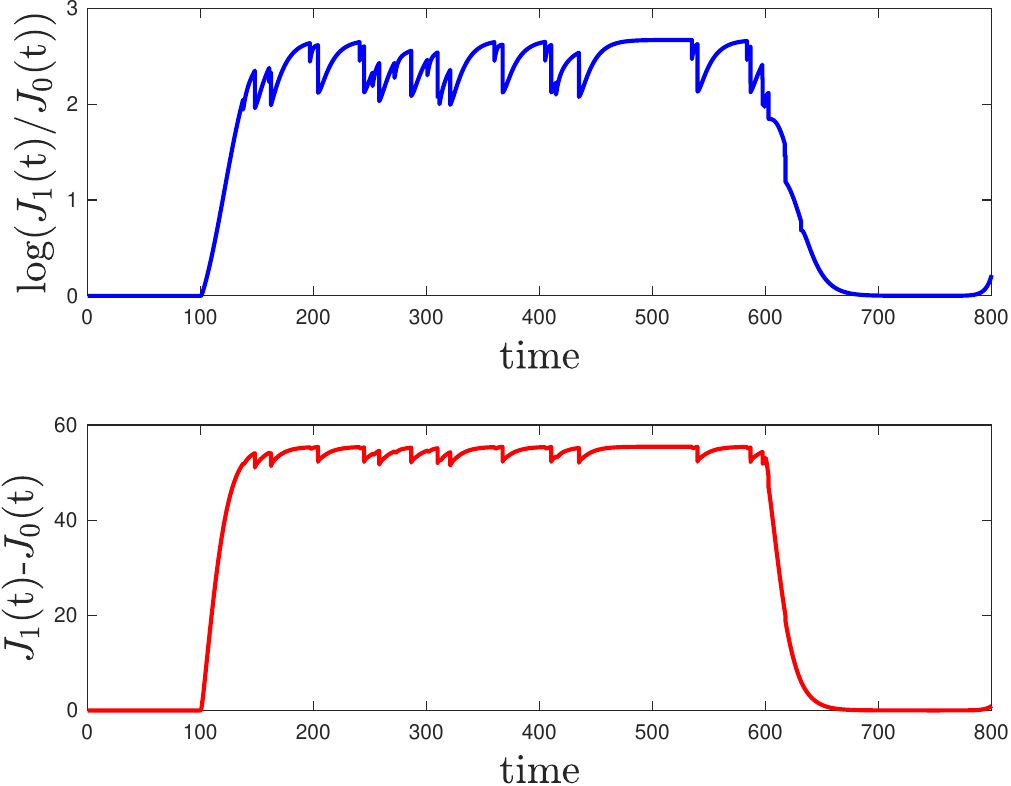}
    \caption{}
    \label{fig:LLRdemo_Js}
    \end{subfigure}

    \caption{(a) Sample persistent $S(t)$ (top plot) and transient $S(t)$ (bottom plot). (b) The log-likelihood ratio for a long signal (or persistent signal) and a short signal. (c) The top and bottom plots show $\log\left(\frac{J_1(t)}{J_0(t)} \right)$ and $J_1(t) - J_0(t)$ in \eqref{eq:L}.}
    \label{fig:sec:llr}        
\end{figure*} 

The aim of this example is to show that we can use the log-likelihood ratio computed by \eqref{eq:L} to distinguish a persistent signal from a transient one. In order to conduct the numerical study, we will assume that the inducer $S$ is produced and degraded by the following chemical reactions:
\begin{subequations}
\label{cr:S_prod} 
\begin{align}
\cee{
S_{\rm pre} &  ->[f_+] S_{\rm pre} +  S \label{cr:S_on}  \\
S &  ->[f_-] \phi \label{cr:S_off} }
\end{align}
\end{subequations}
where \cee{S_{\rm pre}} is a precursor that produces \cee{S}, and $f_+$ and $f_-$ are reaction rate constants. The reason why we choose to use these reactions is that they allow us to use the time profile $S_{\rm pre}(t)$, which is the number of precursor molecules a time $t$, to control the amplitude and duration of $S(t)$. The chemical master equation is used to model the reactions \eqref{cr:S_prod} and \eqref{cr:z_all}. 

We assume that both hypotheses ${\cal H}_0$ and ${\cal H}_1$ include the knowledge of the reactions \eqref{cr:S_prod} and \eqref{cr:z_all} because this knowledge is required for computing ${\rm E}[X_*(t) | {\widetilde{\cal Z}}_{X_*}(t), {\cal H}_i]$ in \eqref{eq:bfiltering}. In order to define the hypotheses, we first define reference signals $R_0(t)$ and $R_1(t)$ for $S_{\rm pre}(t)$. Both reference signals are deterministic ON-OFF pulses and their time profile has the form: 
\begin{eqnarray}
R_i(t) & = & 
\left\{
\begin{array}{ll}
S_{\rm pre}^{\rm ON,ref} & \mbox{for } 0 \leq t < d_i \\
S_{\rm pre}^{\rm OFF,ref} & \mbox{for } t \geq d_i 
\end{array}
\right.
\label{eq:dp:ref} 
\end{eqnarray} 
where $d_i$ is the ON-duration for $R_i(t)$; and, $S_{\rm pre}^{\rm ON,ref}$ and $S_{\rm pre}^{\rm OFF,ref}$, are, respectively, the ON and OFF amplitudes. We assume that $S_{\rm pre}^{\rm ON,ref} \gg S_{\rm pre}^{\rm OFF,ref}$ so that one may consider $S_{\rm pre}^{\rm OFF,ref}$ as a reference value for basal concentration. Furthermore, we assume that the duration of $R_1(t)$ is longer than $R_0(t)$, i.e. $d_1 > d_0$. We define ${\cal H}_0$ (resp.~${\cal H}_1$) as the set of all $S(t)$ that are generated by using $R_0(t)$ ($R_1(t)$) as the time profile for $S_{\rm pre}(t)$. Since $d_1 > d_0$, we will identify ${\cal H}_0$ (resp.~${\cal H}_1$) as, respectively, the set of transient (persistent) signals. 

For this example, the kinetic parameters for the reaction pathway \eqref{cr:z_all} are: $k_+ = 0.02$, $k_- = 0.5$, $g_+ = 0.002$ and $g_-  = 0.05$. The total number of TFs $M$ is 100 and the number of genes $N$ is 1. Furthermore, for reactions \eqref{cr:S_prod}, $f_+ = 0.37$ and $f_- = 0.1$. For the reference signals, we choose $S_{\rm pre}^{\rm ON, ref}$ and $S_{\rm pre}^{\rm OFF, ref}$ to be, respectively, 100 and 3 molecules; and $d_1 = 600$ and $d_0 = 100$. 

We will now use the above set-up to detect a persistent signal. For simplicity, we assume that the actual $S_{\rm pre}(t)$ is the reference signal $R_1(t)$ which results in a persistent input $S(t)$; see the top plot of Fig.~\ref{fig:LLRdemo_St} for a sample of persistent $S(t)$. We first use the chosen $S_{\rm pre}(t)$ and the Stochastic Simulation Algorithm (SSA) \cite{Gillespie:1977ww} to produce a realisation of $\widetilde{Z}_{X_*}(t)$. The simulation assumes that there are zero \cee{S}, \cee{X_*} and $\widetilde{\rm Z}_{X_*}$ molecules initially. We then use optimal Bayesian filtering \cite{Bronstein:2018eh} to numerically compute ${\rm E}[X_*(t) | {\widetilde{\cal Z}}_{X_*}(t), {\cal H}_i]$ for both $i = 0, 1$. The numerical solution to this optimal Bayesian filtering problem requires us to solve a master equation given the hypothesis. In order to solve this master equation exactly, we have purposely chosen the problem parameter values so that the number of $S$ and $X$ molecules are limited. After solving the Bayesian filtering problem, we numerically integrate \eqref{eq:L} to obtain the log-likelihood ratio $L(t)$, which is plotted as the solid blue line in Fig.~\ref{fig:LLRdemo}. We see that the log-likelihood ratio is zero for $t \leq 100$, ramps up in the time interval $[100,600]$ and plateaus after $t \geq 600$. The log-likelihood ratio reaches a positive value at the end, which means correct detection because it says the $\widetilde{Z}_{X_*}(t)$ signal is more likely to have been generated by a persistent signal. 

Next, we use a transient input. We assume that the actual $S_{\rm pre}(t)$ is the reference signal $R_0(t)$; see the bottom plot of Fig.~\ref{fig:LLRdemo_St} for a sample of transient $S(t)$. We perform the same steps as before, namely SSA simulation, optimal Bayesian filtering and numerical integration to obtain the log-likelihood ratio for this transient input. The resulting log-likelihood ratio $L(t)$ is plotted as red dashed lines in Fig.~\ref{fig:LLRdemo}. This $L(t)$ becomes negative which means correct detection. 

Fig.~\ref{fig:LLRdemo_Js} shows the weighting factors $\log\left(\frac{J_1(t)}{J_0(t)} \right)$ and $J_1(t) - J_0(t)$ in \eqref{eq:L} for the case when the $S(t)$ is persistent. (The curves are similar when $S(t)$ is transient.) It shows that these two weighting factors are mostly positive in the time interval [100,600] but are zero outside. This means the contribution to the log-likelihood ratio comes from the signal within $[100,600]$. This can also be seen from Fig.~\ref{fig:LLRdemo} where the log-likelihood ratio does not change outside of [100,600] but increases (resp. decreases) for persistent (transient) signal within [100,600]. This makes intuitive sense because the persistent input is different from the transient input within this time interval, so the signal in this time interval is useful for discriminating persistent signals from transient ones.  

If a persistent signal can give a large positive log-likelihood ratio, then the probability of correctly detecting the persistent signal is higher. For this example, the positive contribution to log-likelihood ratio comes from the first term on the RHS of \eqref{eq:L} because the weighting factors are non-negative, see Fig.~\ref{fig:LLRdemo_Js}. In fact, each time when a \cee{X_*} binds to a $\widetilde{\rm Z}$ in the time interval [100,600], it creates a positive jump in the magnitude of the log-likelihood ratio, which can be seen in Fig.~\ref{fig:LLRdemo}. This means that a persistent signal becomes easier to detect if \cee{X_*} binds to $\widetilde{\rm Z}$ many times when the signal is ON. This can be achieved if the ON duration of the persistent signal has a longer time-scale compared to those of the binding and unbinding reactions of $\widetilde{\rm Z}_{X_*}$ (i.e. reactions \eqref{cr:on2} and \eqref{cr:off2}) so that these reactions occur many times when the input is ON.  

Although this example shows that the solution \eqref{eq:L} can distinguish a persistent signal from a transient one by computing the log-likelihood ratio, the solution is hard to implement in a biochemical environment because Bayesian filtering requires extensive computation and a model of the pathway. In the next section, we will explore how we can compute the log-likelihood ratio approximately without using Bayesian filtering.

%\begin{remark}
%\label{re:properH}
%\revisedtext{
%We mentioned in Sec.~{\ref{sec:dp:sol}} that the hypotheses ${\cal H}_i$ must be properly chosen so that $\log\left(\frac{J_1(t)}{J_0(t)} \right)$ is well defined.  }
%\end{remark}

\section{Computing the log-likelihood ratio approximately} 
\label{sec:approx_llr}
Our ultimate goal is to show that the computation of the log-likelihood ratio $L(t)$ in \eqref{eq:L} can be carried out by a C1-FFL in \eqref{eq:ffl_all}, i.e. there exists a set of parameters for the C1-FFL such that $z(t)$ in \eqref{eq:ffl_all} is approximately equal to $L(t)$ in \eqref{eq:L}. It is not obvious from the expression of \eqref{eq:L} that this can be done. The aim of this section is to derive an ODE, which will be referred to as the {\bf intermediate approximation}, such that the output of this ODE is approximately equal to $L(t)$ when the input is persistent. We will then use this intermediate approximation in Sec.~\ref{sec:c1ffl} to relate to the C1-FFL. 

\subsection{Assumptions}
\label{sec:assumptions}
The detection problem and its solution in Sec.~\ref{sec:results} are general in the sense that they apply to any reaction pathways of the form \eqref{cr:z_all} and hypotheses ${\cal H}_i$. In order to connect the detection problem to the C1-FFL model in \eqref{eq:ffl_all}, we will need to make specific assumptions to derive the intermediate approximation. We will specify these assumptions in this subsection. 

We make the following two assumptions on the reaction pathway \eqref{cr:z_all}: 
\begin{itemize}
\item The time-scale of the inducer-TF reactions \eqref{cr:on1} and \eqref{cr:off1} is faster than that of the TF-gene promoter reactions \eqref{cr:on2} and \eqref{cr:off2}. This translates to small $g_+$ and $g_-$. 
\item The number of TF molecules $M$ is much higher than the number of genes $N$. 
\end{itemize} 

We believe these are realistic assumptions. First, according to \cite[Table 2.2]{Alon}, for {\sl E. coli}, the time-scale for equilibrium binding of small molecules to protein is of the order of 1 ms and the time-scale for TF binding to gene promoter is of the order of 1s. Second, the copy number of most genes is either 1 or 2. 

In order to make the derivation of the intermediate approximation analytically tractable, we assume that hypotheses ${\cal H}_i$ are defined in the same way as in the numerical example in Sec.~\ref{sec:exact:example}. This is so that we can approximate some signal, e.g.~the weighting functions in Fig.~\ref{fig:LLRdemo_Js}, by a piecewise constant function.

For analytical tractability, we further assume that the actual $S_{\rm pre}(t)$ is a deterministic ON-OFF pulse of the form:
\begin{eqnarray}
S_{\rm pre}(t) = 
\left\{
\begin{array}{ll}
\frac{f_-}{f_+} \alpha & \mbox{for } 0 \leq t < d_i \\
S_{\rm pre}^{\rm OFF,ref} & \mbox{for } t \geq d_i
\end{array}
\right.
\label{eq:S_pre_t} 
\end{eqnarray} 
where $d$ is the pulse duration and the ON-amplitude $\frac{f_-}{f_+} \alpha$ will result in a steady state ${\rm E}[S(t)]$ of amplitude $\alpha$ when the pulse is ON. 

We assume that the input signal $S(t)$ is produced by using $S_{\rm pre}(t)$ with the reactions \eqref{cr:S_prod}, and the reaction rate constants $f_+$ and $f_-$ have been chosen such that the signal $S(t)$ is slowly time varying compared to the time-scale of the inducer-TF reactions \eqref{cr:on1} and \eqref{cr:off1}. This makes intuitive sense because a persistent or long $S(t)$ needs to be ``measured" by faster reactions. 

We assume that if a persistent input is applied to the reaction pathway \eqref{cr:z_all}, the pathway is almost at steady state by $d_0$ where $d_0$ is the duration of the reference signal $R_0(t)$ in \eqref{eq:dp:ref}. This requirement can be met if the duration $d_0$ is long enough. It may be instructive to recall from the discussion in the numerical example in Sec.~\ref{sec:exact:example} that there is a time interval which is informative for persistence detection. For the assumptions in this section, the informative time interval can be shown to be $[d_0, \min(d,d_1)]$. Intuitively, this assumption allow us to use the steady state statistics in the time interval $[d_0, \min(d,d_1)]$ to replace $\left[ \frac{d\widetilde{Z}_{X_*}(t)}{dt} \right]_+$ and $g_+ (N - \widetilde{Z}_{X_*}(t))$ in \eqref{eq:L} by, respectively, $g_- \widetilde{Z}_{X_*}(t)$ and  $g_- \frac{\widetilde{Z}_{X_*}(t)}{X_*(t)}$. % Note that we have on purpose put double quotes around the word replace to alert the reader to the fact that the replacement expressions are only heuristically, not mathematically, equivalent. 

\subsection{Intermediate approximation} 
\label{sec:int_approx} 
An ideal persistence detector has the properties that a transient input will result in a zero output and a persistent input will result in a positive output \cite{Alon}. The C1-FFL, when acting as a persistence detector, can be considered to be an approximation of this ideal behaviour \cite{Alon}. However, it is not possible to map the log-likelihood ratio detector in \eqref{eq:L} to the C1-FFL because the log-likelihood ratio becomes negative for transient signals but the concentration in the C1-FFL can only be non-negative. We will use the intermediate approximation as a bridge to connect \eqref{eq:L} to the C1-FFL. The intermediate approximation has two key properties. First, if the input is transient, then the output of the intermediate approximation is zero. Second, if the input is persistent, then the output of the intermediate approximation is approximately equal to the log-likelihood ratio given by \eqref{eq:L}. Another purpose of the intermediate approximation is to replace the complex computation in \eqref{eq:L}, e.g. derivative and optimal Bayesian filtering, by simpler computation that can be implemented by chemical reactions. 

The derivation of the intermediate approximation is given in Appendix \ref{app:ia}, using the assumptions stated in Sec.~\ref{sec:assumptions}. The derivation shows that the time evolution of the intermediate approximation $\hat{L}(t)$ is given by the following ODE: 
\begin{align}
\frac{d\hat{L}(t)}{dt} =& \; \widetilde{Z}_{X_*}(t) \; g_- \; \pi(t) \;  [\phi(X_*(t))]_+  \label{eq:Lfinal} 
\end{align}
where 
\begin{align} 
\phi(X_*(t)) =& \log\left(\frac{X_1}{X_0} \right)  -  \frac{ X_1 - X_0 }{X_*(t)},    \label{eq:phi} \\
X_i =& \frac{M k_+ a_i}{k_+ a_i + k_-} \mbox{ for } i = 0, 1 \label{eq:ia:Xi} \\  
a_0 =& \frac{f_+}{f_-} S_{\rm pre}^{\rm OFF, ref} \label{eq:a_0} \\
a_1 =& \frac{f_+}{f_-} S_{\rm pre}^{\rm ON, ref} \label{eq:a_1} \\
\pi(t) =& \left\{
\begin{array}{cl}
1 & \mbox{for } d_0 \leq t < d_1 \\
0 & \mbox{otherwise}
\end{array}
\right.  \label{eq:pi} \\
\hat{L}(0) &= 0 \label{eq:ia:init}
\end{align}

Furthermore, it can be shown that time evolution of ${\rm E}[\hat{L}(t)]$ obeys the following ODE: 
\begin{eqnarray}
\frac{d{\rm E}[\hat{L}(t)]}{dt} &=& \; {\rm E}[\widetilde{Z}_{X_*}(t)] \; g_- \; \pi(t) \;  [\phi({\rm E}[X_*(t)])]_+ \label{eq:Lfinal_mean} 
\end{eqnarray}

The behaviour of the intermediate approximation $\hat{L}(t)$ depends on two parameters of the input signal $S(t)$: its mean ON-amplitude $\alpha$ and duration $d$ (see \eqref{eq:S_pre_t}). Three important properties for $\hat{L}(t)$ are:
\begin{enumerate}
\item If $d < d_0$, then for all $t$, we have $\hat{L}(t)$ and ${\rm E}[\hat{L}(t)]$ are zero or small. This is due to $\pi(t)$, which is zero outside of $[d_0,d_1)$, and the fact that $X_*(t)$ is likely to be small for $t \geq d_0$. 
\item If the amplitude $\alpha$ is lower than a threshold, then for all $t$, we have $\hat{L}(t)$ is zero or small and ${\rm E}[\hat{L}(t)]$ is zero. We will explain this for ${\rm E}[\hat{L}(t)]$. Since ${\rm E}[X_*(t)]$ is an increasing function of $\alpha$, this means a small $\alpha$ will give a small ${\rm E}[X_*(t)]$. If ${\rm E}[X_*(t)]$ is less than $\frac{X_1 - X_0}{\log\left(\frac{X_1}{X_0} \right)}$ for all $t$, then $ [\phi({\rm E}[X_*(t)])]_+$ on the RHS of \eqref{eq:Lfinal_mean} is zero and this implies ${\rm E}[\hat{L}(t)]$ is zero for all $t$. The explanation for $\hat{L}(t)$ is similar. 
\item If $d$ is longer than $d_0$ and $\alpha$ is sufficiently large, then for $0 \leq t < \min\{d,d_1\}$ we have $\hat{L}(t) \approx L(t)$ where $L(t)$ is given in \eqref{eq:L}. 
\end{enumerate} 

The first two properties are concerned with transient signals, which are those input signals whose duration is no longer than $d_0$ or whose amplitude $\alpha$ is small. The intermediate approximation says that transient signals give a small $\hat{L}(t)$. On the other hand, persistence signals have a duration longer than $d_0$ and have a sufficiently large amplitude $\alpha$. For persistent signals, the intermediate approximation $\hat{L}(t)$ is approximately equal to the log-likelihood ratio $L(t)$ in the time interval $0 \leq t < \min\{d,d_1\}$. Note in particular that this approximation holds for a range of $d$ and $\alpha$ values. From now on, we will choose $d_1$ to be $\infty$ so that $\hat{L}(t) \approx L(t)$ holds for $0 \leq t < d$, i.e.~when the persistent signal is ON.  Note that an infinite $d_1$ means $\pi(t)$ in \eqref{eq:pi} becomes a step function which changes from 0 to 1 at time $d_0$. 

% Comparing to the exact computation of log-likelihood in \eqref{eq:L}, the intermediate approximation can be calculated without using optimal Bayesian filtering. It also removes the need of requiring the knowledge of the pathway \eqref{cr:z_all}. The last point may not be apparent because the constants $f_+$, $f_-$, $S_{\rm pre}^{\rm ON, ref}$, $S_{\rm pre}^{\rm OFF, ref}$ still appear in \eqref{eq:a_0} and \eqref{eq:a_1}. However, one can calculate the intermediate approximation by assuming that the prior knowledge on the values of $a_0$ and $a_1$ (which are, respectively, the steady state values of ${\rm E}[S(t)]$ when the signal is ON and OFF) are available. In other words, one can calculate the intermediate approximation by using the prior information on the mean input ${\rm E}[S(t)]$ as the starting point. 

\subsubsection{Numerical examples} 
\label{sec:int_approx:numer}
% Directory /Users/ctchou/Documents/Work/research/nano/matlab/cffl_pr
% File: plot_paper_ia_mae.m, plot_paper_Xtotal_d.m 0.45\textwidth
\begin{figure}[t]
    \centering
    \begin{subfigure}[t]{0.32\textwidth}
        \centering
        \includegraphics[scale=0.28]{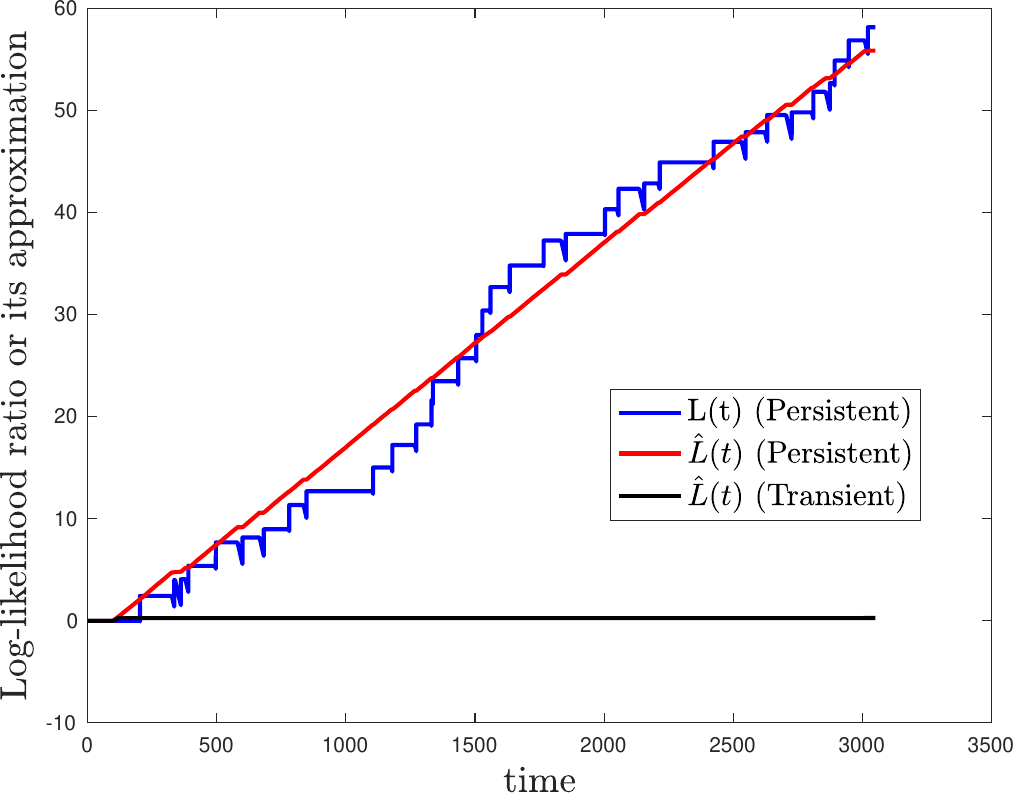}
        \caption{}
        \label{fig:ia_1}
    \end{subfigure} 
        \begin{subfigure}[t]{0.32\textwidth}
        \centering
        \includegraphics[scale=0.28]{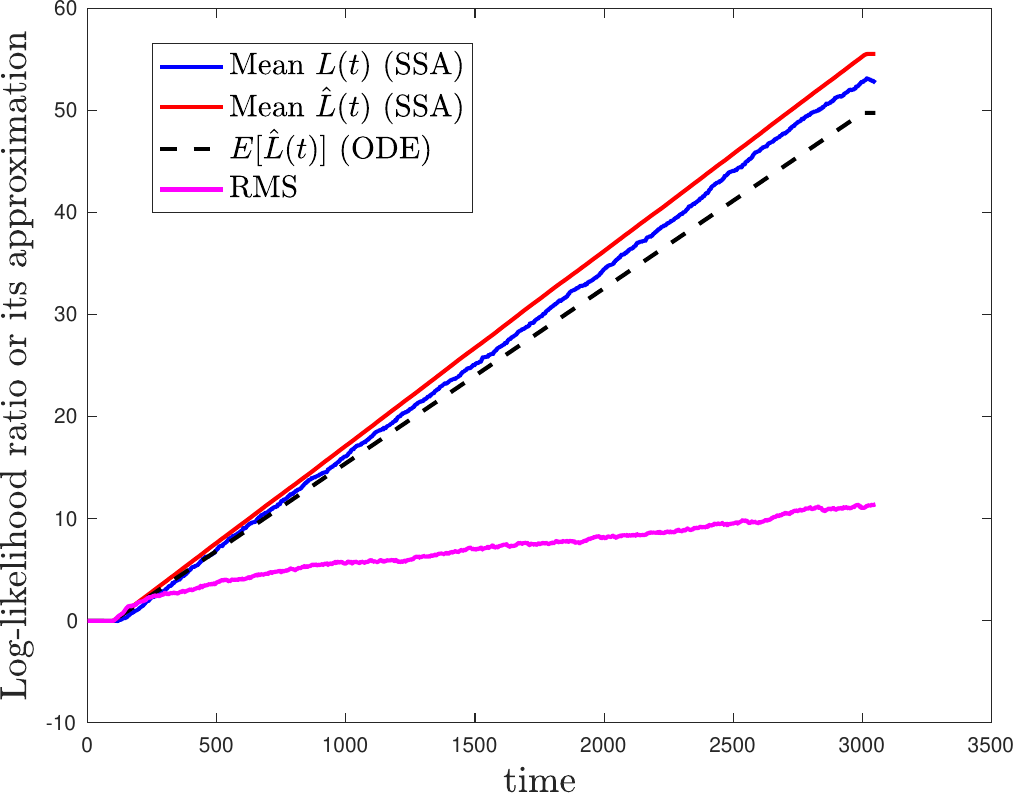}
        \caption{}
        \label{fig:ia_rmse}
    \end{subfigure}     
     \begin{subfigure}[t]{0.32\textwidth}
        \centering
        \includegraphics[scale=0.28]{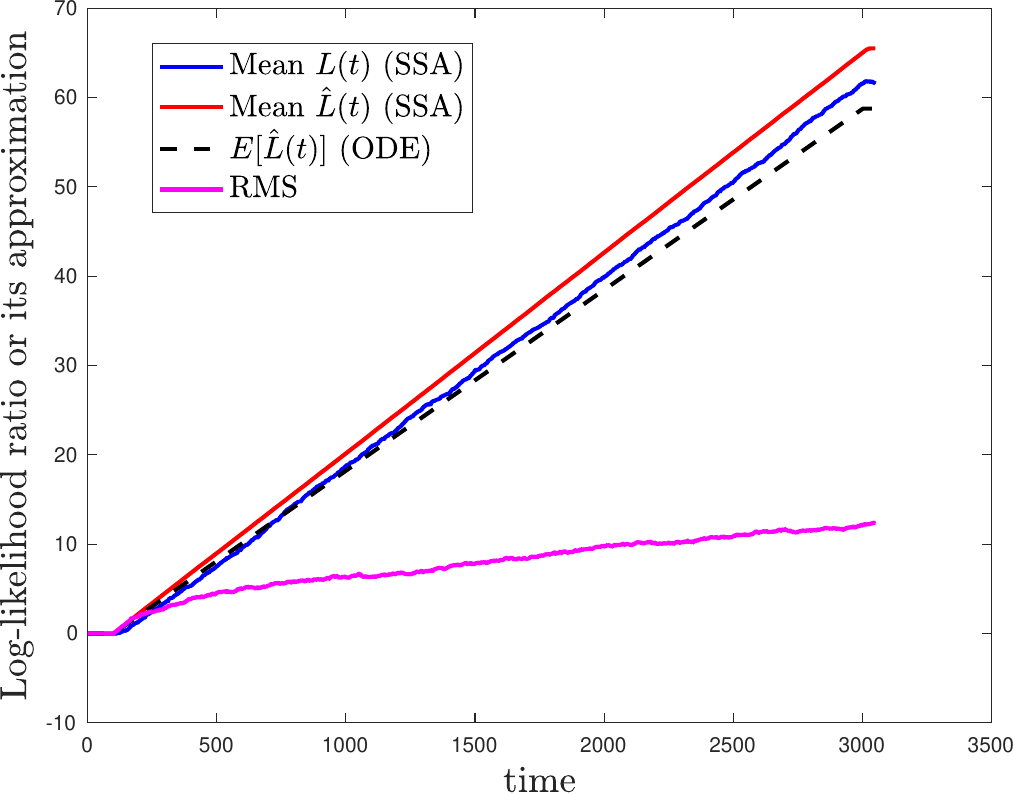}
        \caption{}
        \label{fig:ia_2}
    \end{subfigure}   
\caption{Numerical results for Sec.~\ref{sec:int_approx:numer}. (a) Comparing one realisation of $L(t)$ and $\hat{L}(t)$ for a persistent signal. Also shows a realisation of $\hat{L}(t)$ for a transient signal. (b) Comparing the mean of $L(t)$ from SSA, mean of $\hat{L}(t)$ from SSA, mean of $\hat{L}(t)$ by \eqref{eq:Lfinal_mean}. Also shows RMS error between $L(t)$ and $\hat{L}(t)$ from SSA. (c) Same type of comparison as (b) but for different values of $\alpha$ and $d$.}
\end{figure}

The numerical examples in this section use the following kinetic parameters for the reaction pathway \eqref{cr:z_all}: $k_+ = 0.02$, $k_- = 0.5$, $g_+ = 0.002$ and $g_- = 0.0125$. These parameters have been chosen such that the time-scale of the inducer-TF reactions are faster than those of the TF-gene promoter. The number of TF $M$ is 100 and the number of genes $N$ is 1, which means $M \gg N$. The values of $f_+$, $f_-$, $S_{\rm pre}^{\rm ON,ref}$ and $S_{\rm pre}^{\rm OFF,ref}$ are the same as those in Sec.~\ref{sec:exact:example}. The duration parameters for the reference signals are: $d_0 = 100$, $d_1 = \infty$. The signal duration $d = 3000$. All the above parameters are fixed. We will vary the values of amplitude $\alpha$. 

For the first numerical experiment, we use $\alpha = 37$, which means $S_{\rm pre}(t)$ has an amplitude of $S_{\rm pre}^{\rm ON,ref}$ when it is ON. This excitation means there is a mean probability of 0.6 that \cee{X} is active. We use SSA simulation to generate 100 realisations of $X_{*}(t)$ and $\widetilde{Z}_{X_*}(t)$, and use them to compute the true log-likelihood ratio $L(t)$ (which requires only $\widetilde{Z}_{X_*}(t)$) and the intermediate approximation $\hat{L}(t)$ (which requires both $X_{*}(t)$ and $\widetilde{Z}_{X_*}(t)$). Fig.~\ref{fig:ia_1} compares one realisation of $L(t)$ (blue line) and $\hat{L}(t)$ (red line) for a persistent input. It can be seen that the intermediate approximation is fairly accurate. 

We then use all 100 realisations to compute the mean of $L(t)$, the mean of $\hat{L}(t)$, and the root-mean-square (RMS) error of $L(t) - \hat{L}(t)$. The results are plotted in Fig.~\ref{fig:ia_rmse}. This shows that $\hat{L}(t)$ is a good approximation of $L(t)$ for many realisations. 
% The results also show that $\hat{L}(t)$ is a slightly bias estimate of $L(t)$. 
% The RMS error looks large compared to the bias because $L(t)$ is a much noisier signal compared with $\hat{L}(t)$, e.g.~the mean and standard deviation for $L(t)$ at $t = d$ is $45.03 \pm 9.52$, while those for $\hat{L}(d)$ are $51.47 \pm 3.41$. 
Next, we check the accuracy of using \eqref{eq:Lfinal_mean} to compute ${\rm E}[\hat{L}(t)]$. The black dashed lines in Fig.~\ref{fig:ia_rmse} plot ${\rm E}[\hat{L}(t)]$ computed from \eqref{eq:Lfinal_mean}, which is almost the same as the mean obtained via SSA. This shows that \eqref{eq:Lfinal_mean} is an accurate method to compute ${\rm E}[\hat{L}(t)]$. 

The discussion so far focused on persistent signals. The black line in Fig.~\ref{fig:ia_1} shows the intermediate approximation for a transient input with $d = 100$. It can be seen that the output is small. This agrees with our earlier prediction.

For a given set of hypotheses, the intermediate approximation holds for a range of  $\alpha$ and $d$. We now change $\alpha$ to 99.9 (which means $S_{\rm pre}(t)$ has an amplitude of 1.33$S_{\rm pre}^{\rm ON,ref}$ when it is ON). By using 100 rounds of SSA simulations, we compute the the means of both $L(t)$ and $\hat{L}(t)$, as well as the RMS error between them. The results are in Fig.~\ref{fig:ia_2} and they show that the intermediate approximation is accurate for different values of $\alpha$. Fig.~\ref{fig:ia_2} also shows ${\rm E}[\hat{L}(t)]$ computed from \eqref{eq:Lfinal_mean} and we can see that the accuracy is good. 

\begin{remark}
The reader may wonder why we do not define the hypotheses of the detection problem as: ${\cal H}_0$ (resp. ${\cal H}_1$) means the duration of the input signal is shorter (longer) than a given threshold.  The reason is that these are composite hypotheses and the solution to the resulting detection problem is much harder, see \cite[Remark 5.1]{Chou:2018jh} for a more in-depth discussion. We also want to point out that, even with our simpler formulation, the resulting detector gives a small output for any signal whose duration is shorter than $d_0$. 
\end{remark}

\ifarxiv
\subsection{Relevance to biological detection problems} 
\label{sec:bio_detection}
In this section, we will discuss why the likelihood ratio is a relevant criterion for some biological detection problems. For this exposition, we will assume that the purpose of the detection is to determine if a certain nutrient is persistently present and if yes, then the organism wants to produce the enzyme to consume the nutrient. Our discussion is based on Bayesian decision theory which considers both utility and cost of actions, e.g. the successful detection and consumption of a nutrient gives a positive utility to the organism at the cost of producing the enzyme. 

We will use the hypotheses ${\cal H}_0$ and ${\cal H}_1$ to refer to the conditions that the nutrient is respectively, absent and present, in the environment. We will also use negative and positive, respectively, to refer to the two hypotheses ${\cal H}_0$ and ${\cal H}_1$. Given these two hypotheses, the detection problem may decide for either ${\cal H}_0$ (negative) or ${\cal H}_1$ (positive). Table \ref{tab:why_lr} summarises the four possible combinations from the two environmental conditions and the two detection outcomes. These four combinations are labelled as True Negative (TN), False Negative (FN), False Positive (FP) and True Positive (TP) using the terminologies commonly used in statistics. For the time being, we assume that the utility can only take two values: $U_1$ and $U_0$ where $U_1 > U_0 = 0$. The living organism can only get the positive utility $U_1$ for TP as this is the only situation which the nutrient is present and detected. Similarly, for the cost, we assume $C_1 > C_0 = 0$. The living organism incurs a cost when FP or TP occurs as enzymes are made in these cases. 

Let $P_0$ and $P_1$ denote the true probabilities that the nutrient is, respectively, absent and present. Let also $P_{TP}$ denote the probability of TP etc. The mean utility is $P_{TP} P_1 U_1$ and the mean cost is $P_{FP} P_0 C_1 + P_{TP}P_1 C_1$. Let $C_{\rm max}$ be the maximum cost that the living organism can afford. From a Bayesian decision point of view, the goal is to maximise the mean utility subject to a constraint on the mean cost, i.e.
\begin{eqnarray}
& & \max P_{TP} P_1 U_1 \label{eq:max_util}  \\
& & \mbox{subject to } P_{FP} P_0 C_1 + P_{TP}P_1 C_1 \leq C_{\rm max} \nonumber 
\end{eqnarray}
We show in Appendix \ref{app:util} that the solution of this utility maximisation problem is to choose a suitable positive threshold $\rho$ such that the organism should decide for ${\cal H}_1$ if the likelihood ratio $\frac{{\rm P}[{\rm data} | {\cal H}_1] }{{\rm P}[ {\rm data} | {\cal H}_0]} \geq \rho$. In fact, the derivation shows that, if $U_1 > \lambda C_1$ where $\lambda$ is the Lagrangian multiplier of the above optimisation problem, then the above utility maximisation problem is equivalent to maximising $P_{TP}$ subject to an upper bound on $P_{FP}$, which is in fact the scenario covered by the Neyman-Pearson lemma. This shows that the likelihood ratio is a suitable statistic to be used. Note that we have assumed for simplicity that $U_0 = 0$ and $C_0 = 0$ earlier, however it can be shown that, under some conditions, the result still holds for non-zero $U_0$ and $C_0$. 

\begin{table}[t]
\centering
\begin{tabular}{| l | l | l | l |} \hline
      \multicolumn{2}{|c|}{}      &    \multicolumn{2}{|c|}{Environmental conditions}    \\ \cline{3-4}
     \multicolumn{2}{|c|}{}        &  Negative & Positive \\ \hline 
\multirow{6}{*}{Detection outcomes}       & \multirow{3}{*}{Negative} &  True Negative (TN)  & False Negative (FN) \\
       &  &  Utility = $U_0$ & Utility = $U_0$ \\ 
       &  &  Cost  = $C_0$ & Cost = $C_0$ \\         \cline{2-4}
      & \multirow{3}{*}{Positive} & False Positive (FP) & True Positive (TP)  \\ 
             &  &  Utility = $U_0$  & Utility = $U_1$ \\ 
       &  &  Cost = $C_1$ & Cost = $C_1$ \\         \hline  
\end{tabular} 
\caption{The utilities and costs of the four combinations of environmental conditions and detection outcomes.}
\label{tab:why_lr}
\end{table}
\fi

\begin{figure}[t]
\centering
\includegraphics[scale=0.29]{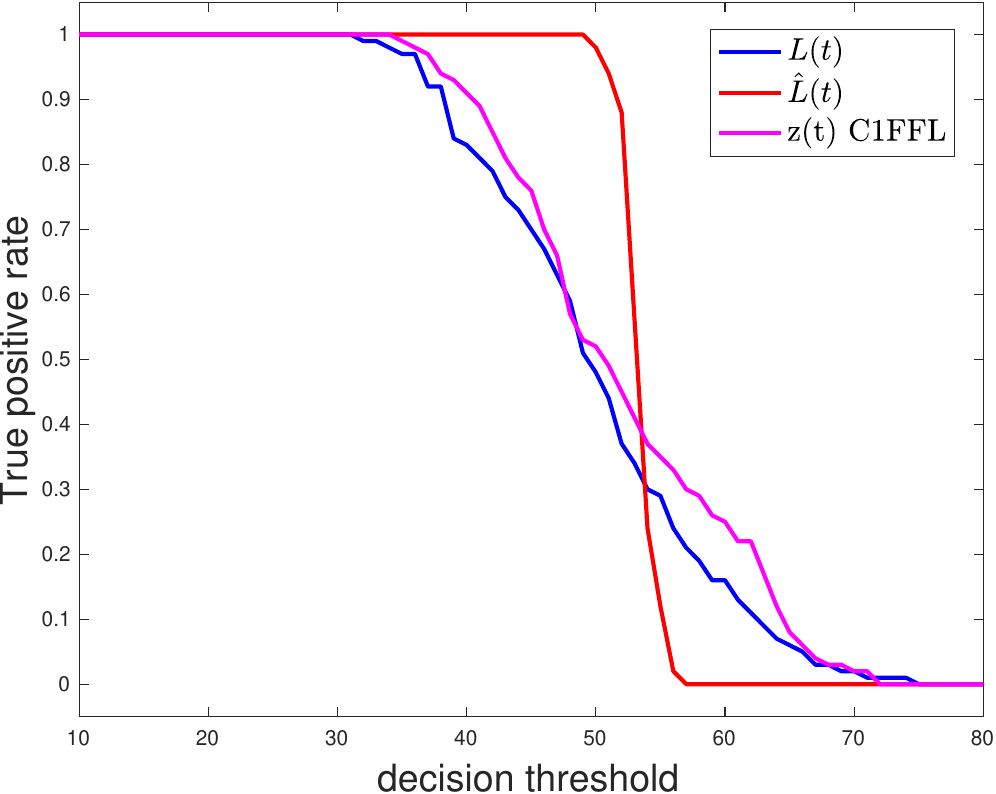}    
\caption{The impact of the decision threshold on the TP rates.}
\label{fig:tp}
\end{figure}

\ifarxiv
 \subsubsection{Numerical example} 
\label{sec:int_approx:tpfp}
\else
\subsubsection{True positive (TP) and false positive (FP) rates} 
\label{sec:int_approx:tpfp}
\fi
In this section, we want to study the performance of using the approximate log-likelihood ratio $\hat{L}(t)$ for decision making. Since our interest is to detect persistent signals, we can do that by testing whether the log-likelihood ratio is greater than or equal to a positive threshold $\rho$. 
We use a numerical example to illustrate the impact of the positive decision threshold $\rho$ on the TP and FP rates.
For the TP rate, we assume the input is a long pulse of duration $d = 3000$ and generate 100 realisations of $X_{*}(t)$ and $\widetilde{Z}_{X_*}(t)$. We use the data to compute 100 values of $L(d)$ and $\hat{L}(d)$ where $L(d)$ and $\hat{L}(d)$ are, respectively, the log-likelihood ratio \eqref{eq:L} and approximate log-likelihood ratio \eqref{eq:Lfinal} at time $t = d$.  We then compare these values against the threshold to obtain the TP rates. We vary the threshold between 10 and 80. The blue and red lines in Fig.~\ref{fig:tp} show the TP rates for $L(t)$ and $\hat{L}(t)$ respectively. It shows that there is a wider range of thresholds that can be used to obtain a high TP rate for $\hat{L}(t)$. This is because $\hat{L}(t)$ has a lower variance in comparison. We will discuss the magenta curve in Fig.~\ref{fig:tp} in Sec.~\ref{sec:c1ffl:fit:ex}. 

For the FP rate, we generate 100 realisations using a short pulse of duration $d = 100 (= d_0)$ and use the data to compute $L(d)$ and $\hat{L}(d)$. For all the 100 realisations, the $L(d)$'s are negative and $\hat{L}(t) \approx 0$. Therefore for all the thresholds in $10-80$, the FP rates are zero for both $\hat{L}(t)$ and $L(t)$. This shows that we are able to find thresholds that give a large TF while keeping FP low. 

The above numerical experiment on computing the FP rate also shows that it is not a problem to ``round" the negative log-likelihood ratio in $L(d)$ to a near zero approximation $\hat{L}(d)$. This is because, if the test is to check whether the log-likelihood ratio is above a sufficiently positive threshold, then a negative $L(d)$ or an almost zero $\hat{L}(d)$ will lead to the same decision. 

%\subsection{Discussion}
%\revisedtext{
%In order to derive the intermediate approximation, we make the assumption that input signal $S(t)$ is produced by using the reactions {\eqref{cr:S_prod}}. An open problem is whether the intermediate approximation still holds for other reactions for producing $S(t)$. 
%}

\section{Using the C1-FFL to approximately compute log-likelihood ratio}
\label{sec:c1ffl}

We have shown in the previous section that the mean of the intermediate approximation ${\rm E}[\hat{L}(t)]$ is an accurate approximation of the mean log-likelihood ratio ${\rm E}[L(t)]$ when the input is persistent. The aim of this section is to show that we can use the C1-FFL in \eqref{eq:ffl_all} to approximately compute ${\rm E}[\hat{L}(t)]$ in \eqref{eq:Lfinal_mean}. 

\subsection{Relating ${\rm E}[\hat{L}(t)]$ to the C1-FFL}  
\label{sec:hatL:c1ffl}

\begin{figure}[t]
   \centering
        \includegraphics[page=4,scale=0.25,trim={0.5cm 0.5cm 5cm 0.5cm},clip]{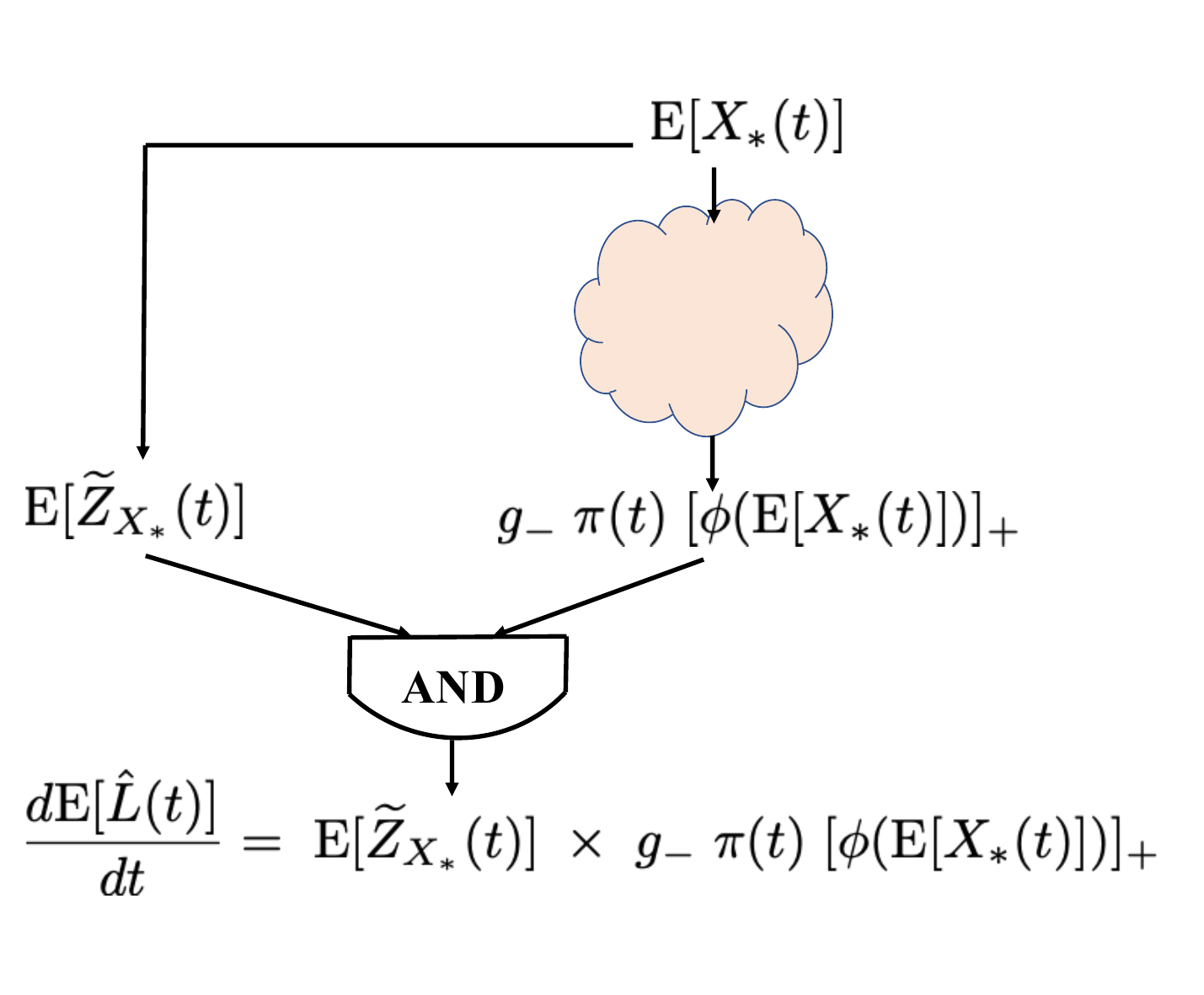}   
        \caption{Relating the computation of the mean approximate log-likelihood ratio ${\rm E}[\hat{L}(t)]$ in \eqref{eq:Lfinal_mean} to the C1-FFL.}
        \label{fig:compute:Lhat}
\end{figure}

For the time being, we will assume $d_z$ in \eqref{eq:ffl3} is zero and show that it is possible to find C1-FFL parameters such that $z(t)$ in \eqref{eq:ffl_all} is approximately equal to ${\rm E}[\hat{L}(t)]$ in \eqref{eq:Lfinal_mean}. We will explain in Remark \ref{re:nonzero_dz} how a non-zero $d_z$ can be handled. 

% We will do this by showing that there exist C1-FFL parameters such that the output $z(t)$ of the C1-FFL is approximately equal to ${\rm E}[\hat{L}(t)]$ for a range of $a$ and $d$ values. 

We depict the calculations of \eqref{eq:Lfinal_mean} in Fig.~\ref{fig:compute:Lhat}. We split the computation on the RHS of \eqref{eq:Lfinal_mean} as the product of $ {\rm E}[\widetilde{Z}_{X_*}(t)]$ and $g_- \; \pi(t) \;  [\phi({\rm E}[X_*(t)])]_+$ where the multiplication operation is depicted as an {\sc and} gate in the figure. Note that the two-branch structure of the computation in Fig.~\ref{fig:compute:Lhat} has a direct resemblance with that of the C1-FFL in Fig.~\ref{fig:c1ffl_b}. We first consider the computation of the two branches separately. 

\begin{figure*}[t]
    \centering
    \begin{subfigure}[t]{0.32\textwidth}
        \centering
        \includegraphics[scale=0.28]{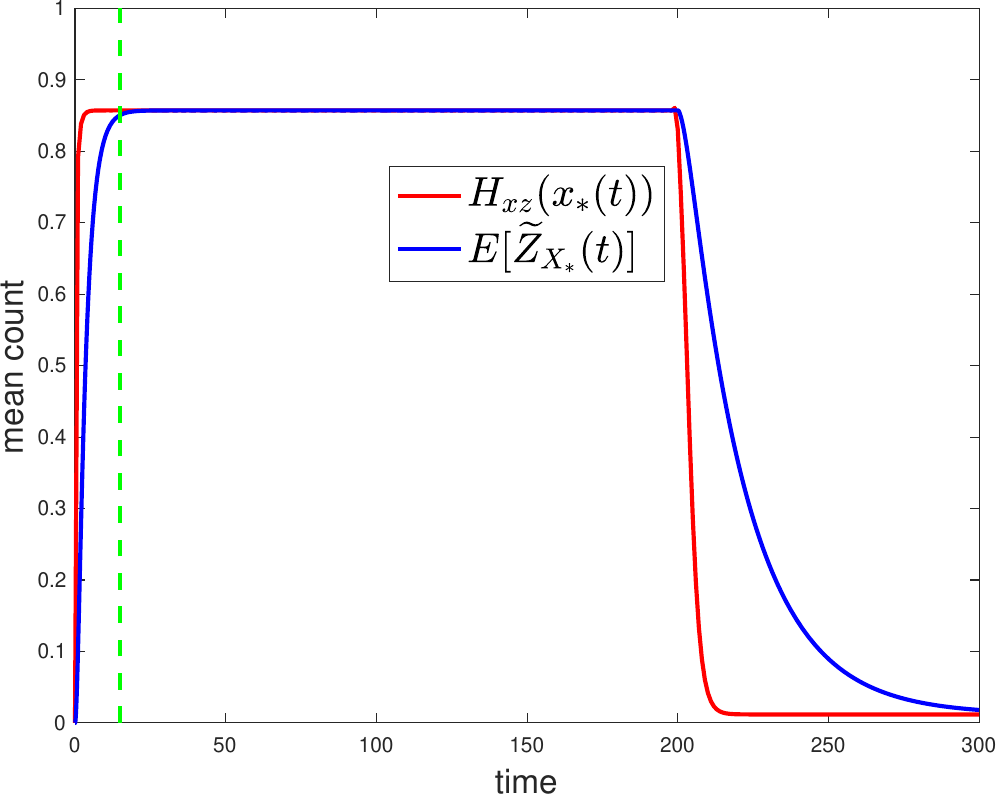}
        \caption{}
        \label{fig:match:c1ff1_1}
    \end{subfigure} 
    \begin{subfigure}[t]{0.32\textwidth}
        \centering
        \includegraphics[scale=0.28]{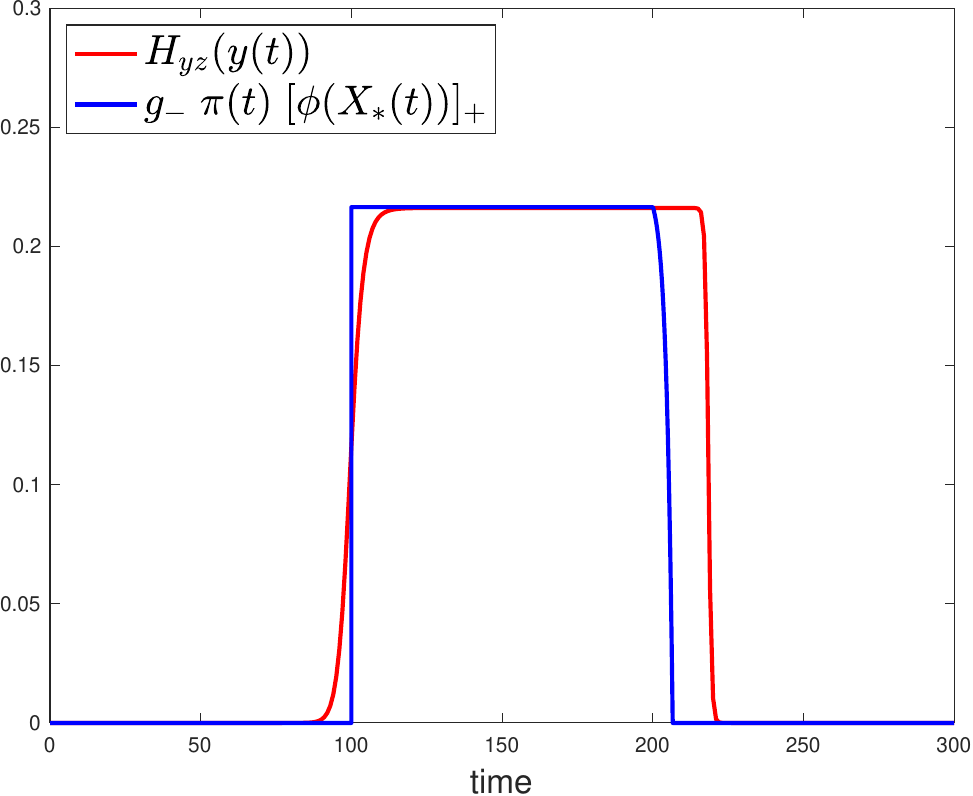}
        \caption{}
        \label{fig:match:c1ff1_2}
    \end{subfigure}       
     \begin{subfigure}[t]{0.32\textwidth}
        \centering
        \includegraphics[scale=0.28]{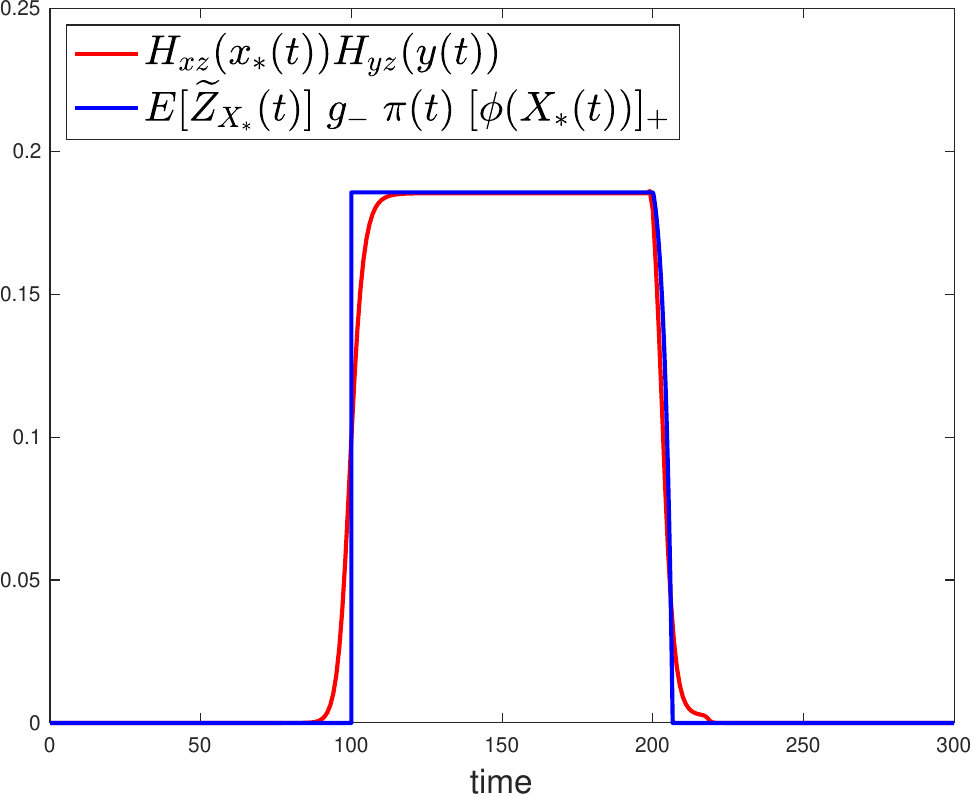}
        \caption{}
        \label{fig:match:c1ff1_3}
    \end{subfigure}       
\caption{(a) Comparing $\widetilde{z}_{X_*}(t)$ and  $\frac{N g_+ x_*(t)}{g_+ x_*(t) + g_-}$. 
(b) Comparing $g_- \; \pi(t) \; [\phi(x_*)]_+$ and $H_{yz}(y(t))$. 
(c) Comparing $\widetilde{z}_{X_*}(t) g_- \; \pi(t) \;  [\phi(x_*)]_+$ and $H_{xz}(x_*(t)) H_{yz}(y(t))$.
}
\label{fig:use_hill}
\end{figure*}

We first consider the computation of ${\rm E}[\widetilde{\rm Z}_{X_*}(t)]$ from ${\rm E}[{\rm X}_*(t)]$, which is the branch on the left in Fig.~\ref{fig:compute:Lhat}. By using the volume scaling assumption (Sec.~\ref{sec:results}), we equate molecular count $\widetilde{Z}_{X_*}(t)$ with concentration $\widetilde{z}_{X_*}(t)$, and similarly, $X_*(t)$ with $x_*(t)$. We propose to compute $\widetilde{z}_{X_*}(t)$ from $x_*(t)$ by using:
\begin{eqnarray}
\widetilde{z}_{X_*}(t)  & =_{\mbox{\scriptsize computed  by}} &  \frac{N g_+ x_*(t)}{g_+ x_*(t) + g_-} \label{eq:map:ZXstar} 
\end{eqnarray} 
Note that the above approximation is obtained from assuming that the TF-gene reactions \eqref{cr:on2} and \eqref{cr:off2} are close to steady state. By using the parameters in the numerical example in Section \ref{sec:int_approx:numer}, we have plotted the two sides of \eqref{eq:map:ZXstar} in Fig.~\ref{fig:match:c1ff1_1} when the input is persistent with a duration $d$ of 200. Note that there are two transients of $\widetilde{z}_{X_*}(t)$ in the figure, one after time 0 when the input turns ON and the other at time $d$ when the input turns OFF. We can see that, other than these two transients, the two sides of \eqref{eq:map:ZXstar} are almost equal. We will show later on these two transients have little effect on the accuracy of the overall computation. Note that the RHS of \eqref{eq:map:ZXstar} has the form of a Hill function and we can identify it with $H_{xz}(x_*(t) )$ in \eqref{eq:ffl3}. 

We next consider the computation of $g_- \; \pi(t) \;  [\phi({\rm E}[X_*(t)])]_+$ from $ {\rm E}[X_*(t)]$, which is depicted by a cloud in Fig.~\ref{fig:compute:Lhat}. We first argue that, for most of the admissible choices of $d_0$, there must be a time delay element in the cloud. This implies that there must be some chemical reactions in the cloud in order to create this time delay. To understand which $d_0$ is admissible, we recall that we assume in Section \ref{sec:assumptions} that the pathway \eqref{cr:z_all} is almost at steady state by the time $d_0$. As an illustration of this assumption, consider Fig.~\ref{fig:match:c1ff1_1} which shows the time profile of $\widetilde{z}_{X_*}(t)$, the vertical dashed line shows the time (which we will denote as $t_{\rm ss}$) by which $\widetilde{z}_{X_*}(t)$ is sufficiently close to steady state, the assumption means an admissible $d_0$ must be greater than or equal to $t_{\rm ss}$. 

\ifarxiv
In Appendix \ref{app:delay}, we show that there must be a delay in the cloud if $d_0$ is strictly greater than $t_{\rm ss}$. 
\else
In our technical report \cite[Appendix D]{Chou:arxiv_ffl2}, we show that there must be a delay in the cloud if $d_0$ is strictly greater than $t_{\rm ss}$.
\fi
Intuitively, a delay is needed because $\pi(t)$ is zero in $[0,d_0)$ but \cee{X_*} almost reaches steady state before $d_0$. Given that there must be a delay element in the cloud in Fig.~\ref{fig:compute:Lhat}, we can achieve that by inserting a transcription node (e.g. Node $Y$ in Fig.~\ref{fig:c1ffl_b}) in the cloud. With this insertion, we can identify the branch on the right in Fig.~\ref{fig:compute:Lhat} with the indirect branch in the C1-FFL in Fig.~\ref{fig:c1ffl_b}. 

The next step is to show that we can find Hill functions $H_{xy}()$ and $H_{yz}()$ in \eqref{eq:ffl_all} which will enable us to compute $g_- \; \pi(t) \;  [\phi(x_*(t))]_+$.  Since the inducer-TF pathway is fast, we can approximate $x_*(t)$ by its steady state value $x_{*,{\rm ss}}$ for $t < d$. We can use \eqref{eq:ffl2} to show that
\begin{eqnarray}
y(t) & = & \frac{H_{xy}(x_{*,{\rm ss}})}{d_y} (1 - \exp(- d_y t)) \label{eq:opt:yt} 
\end{eqnarray}
Our aim is to achieve the approximation: 
\begin{eqnarray}
H_{yz}(y(t)) & \approx &  g_- \; \pi(t) \;  [\phi(x_{*,{\rm ss}})]_+.    \label{eq:opt:match2}  
\end{eqnarray}
We will turn this into two requirements. First, in order to imitate $\pi(t)$, we require $H_{yz}(y(t))$ to have a sharp transition from a low value to a high value around $d_0$. Note that this requirement implies that $y(t)$ should not have reached steady state by $d_0$ and can be achieved by suitable choice of $d_y$. Second, when $H_{yz}(y(t))$ reaches steady state, which happens some time after time $d_0$, its amplitude is determined by $g_- [\phi(x_*)]_+$. We can consider these two requirements separately because the first requirement is concerned with a time event around time $d_0$ while the second requirement is concerned with the steady state amplitude which is reached at a time later than $d_0$. 

We first consider how $H_{yz}(y(t))$ can be used to realised a transition around time $d_0$. Let us recall that $H_{yz}(y(t))$ has the form $\frac{h_{yz} y(t)^{n_{yz}}}{K_{yz}^{n_{yz}} + y(t)^{n_{yz}}}$. Given that $y(t)$ is an increasing function of $t$, if we choose $K_{yz}$ to be $y(d_0)$, then we have $y(t) < K_{yz}$ for $t < d_0$ and $y(t) > K_{yz}$ for $t > d_0$. Hence, if $n_{yz}$ is sufficiently large, then $H_{yz}(y(t))$ will have a sharp transition around $d_0$. 

The argument in the last paragraph works for a particular choice of $x_{*,{\rm ss}}$, which corresponds to a particular choice of input signal amplitude. We will now explain how $H_{yz}(y(t))$ can be used to realised a transition at time $d_0$ for a range of input amplitudes. Since $H_{xy}(x_{*,{\rm ss}})$ is a Hill function, then for sufficiently large $x_{*,{\rm ss}}$, the value of $H_{xy}(x_{*,{\rm ss}})$ does not change much because $H_{xy}(x_{*,{\rm ss}})$ saturates. As a result, there is a range of $x_{*,{\rm ss}}$ such that $y(t)$ does not change a lot. This means if we choose $K_{yz}$ to be the $y(d_0)$ corresponding to a particular $x_{*,{\rm ss}}$ which saturates $H_{xy}(x_{*,{\rm ss}})$, we can a achieve sharp transition around time $d_0$ for a range $x_{*,{\rm ss}}$. 

We will now give an illustration on the first requirement. Fig.~\ref{fig:match:c1ff1_2} plots the two sides of \eqref{eq:opt:match2} for a particular input amplitude. We see that we can use $H_{yz}$ to approximate $g_- \; \pi(t) \;  [\phi(x_{*,{\rm ss}})]_+$ well except near the falling edge of $g_- \; \pi(t) \;  [\phi(x_{*,{\rm ss}})]_+$. We can now consider the two branches in Fig.~\ref{fig:compute:Lhat} together. We now compare the whole RHS of \eqref{eq:Lfinal_mean}, which is ${\rm E}[\widetilde{Z}_{X_*}(t)]$ times $g_- \; \pi(t) \;  [\phi({\rm E}[X_*(t)])]_+$, and the RHS of  \eqref{eq:ffl3}, which is $H_{xz}(x_*(t))$ times $H_{yz}(y(t))$. Fig.~\ref{fig:match:c1ff1_3} compares these two expressions. An interesting observation is that the mismatch in some time intervals in Fig.~\ref{fig:match:c1ff1_1} and \ref{fig:match:c1ff1_2} are cancelled out when the multiplication is made. % Note that this example is for a particular value of $x_*$, we will present an example that works for a range of $x_*$ in Sec.~\ref{sec:c1ffl:fit:ex}. 

We now consider the second requirement which requires the steady state value of $H_{yz}(y(t))$ to be  $g_- [\phi(x_*)]_+$. According to \eqref{eq:opt:yt}, the steady state amplitude of $y(t)$ is $\frac{H_{xy}(x_{*,{\rm ss}})}{d_y}$, therefore the requirement is equivalent to finding $H_{xy}$ and $H_{yz}$ such that:
\begin{eqnarray}
H_{yz} \left( \frac{H_{xy}(x_{*,{\rm ss}})}{d_y} \right) \approx g_- [\phi(x_{*,{\rm ss}})]_+
\label{eq:match:xss} 
\end{eqnarray}  
holds for a large range of $x_{*,{\rm ss}}$. Let $\chi = \frac{1}{x_{*,{\rm ss}}}$. By using the expressions of $H_{xy}()$ and $H_{yz}()$ in \eqref{eq:ffl_all} as well as $\phi()$ in \eqref{eq:phi}, we can rewrite \eqref{eq:match:xss} in terms of $\chi$ as:
\begin{eqnarray}
& \frac{h_{yz}}{\eta (1 + K_{xy}^{n_{xy}} \chi^{n_{xy}})^{n_{yz}}+1} \nonumber \\
& \approx g_- \left[\log\left( \frac{X_1}{X_0} \right) - (X_1 - X_0) \; \chi \right]_+ 
\label{eq:match:chi} 
\end{eqnarray} 
where $\eta = \left( \frac{K_{yz} d_y}{h_{xy}} \right)^{n_{yz}}$. If we consider the expression on the RHS of \eqref{eq:match:chi} as a function of $\chi$, it has a linearly decreasing segment followed by a constant segment at zero. Recall that we have chosen $n_{yz}$ to be large earlier, and if we furthermore choose $n_{xy}$ is to be 1, then the LHS may be approximated by three segments: a constant for small $\chi$, linearly decreasing for intermediate $\chi$ and zero for large $\chi$. This shows that it is possible to find Hill functions $H_{xy}$ and $H_{yz}$ such that the second requirement holds. We will demonstrate this with a numerical example. 

% \eqref{eq:hyz:pi} where we aim to approximate the step function $\pi(t)$ which changes from 0 to 1 at $d_0$. 
% For a given value of $x_*$, \eqref{eq:hyz:phi} shows how $h_{yz}$ can be chosen. 

% We first note that in the above approximation, $y(t)$ is an increasing function of $t$ and $\pi(t)$ is a step. If $H_{yz}()$ is a Hill function of the form in \eqref{eq:ffl3}, we now argue that we can find $h_{yz}$, $n_{yz}$ and $K_{yz}$ such that:
%\begin{eqnarray}
%\frac{y(t)^{n_{yz}}}{K_{yz}^{n_{yz}} + y(t)^{n_{yz}}}  & \approx &  \pi(t) \label{eq:hyz:pi}   \\
%h_{yz} & \approx & g_- \;   [\phi(x_*)]_+ \label{eq:hyz:phi} 
%\end{eqnarray} 

% Second, note that the function $[\phi(x_*)]_+$ flattens out if $x_*$ is sufficiently large, therefore $h_{yz}$ can be chosen so that \eqref{eq:hyz:phi} holds for a range of $x_*$. Overall, this means we can choose the parameters of the Hill functions $H_{xy}$  and $H_{yz}$ so that \eqref{eq:opt:match2} holds for a range of input amplitudes. 

\subsection{Numerical example} 
\label{sec:c1ffl:fit:ex}
% plot_LLR_opt_v6
% study_LLR_opt_v6 
\begin{figure*}[t]
    \centering
    \begin{subfigure}[t]{0.32\textwidth}
        \centering
        \includegraphics[scale=0.28]{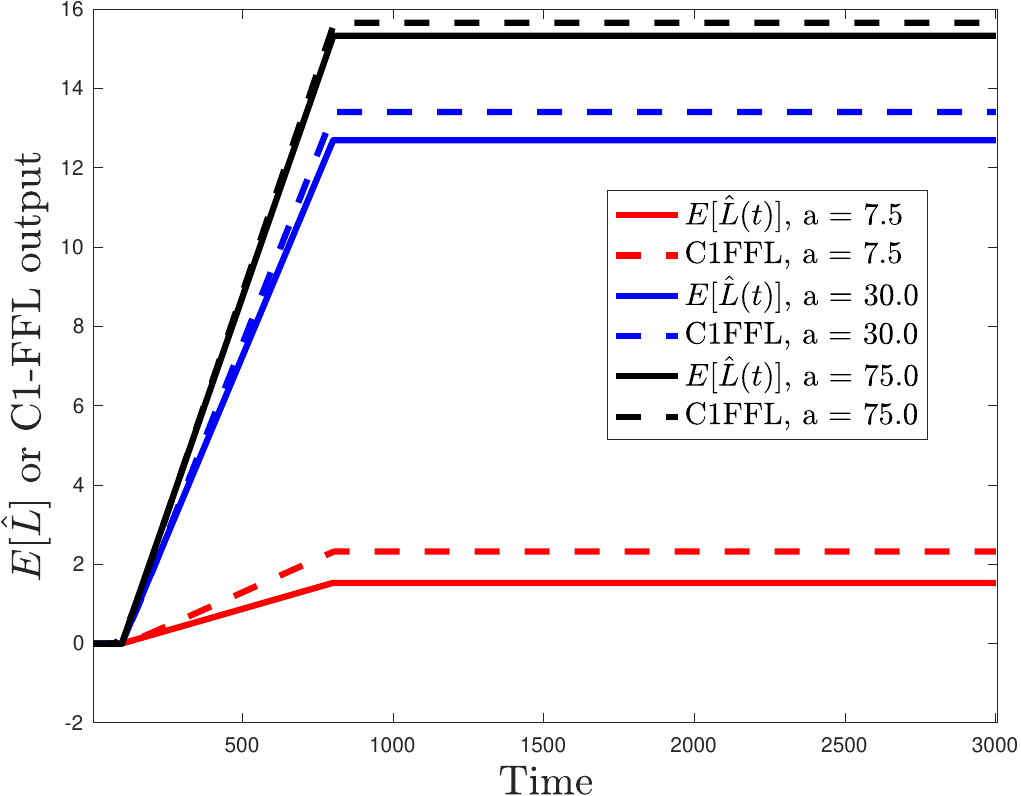}
        \caption{}
        \label{fig:opt_c1ffl}
    \end{subfigure} 
    \begin{subfigure}[t]{0.32\textwidth}
        \centering
        \includegraphics[scale=0.28]{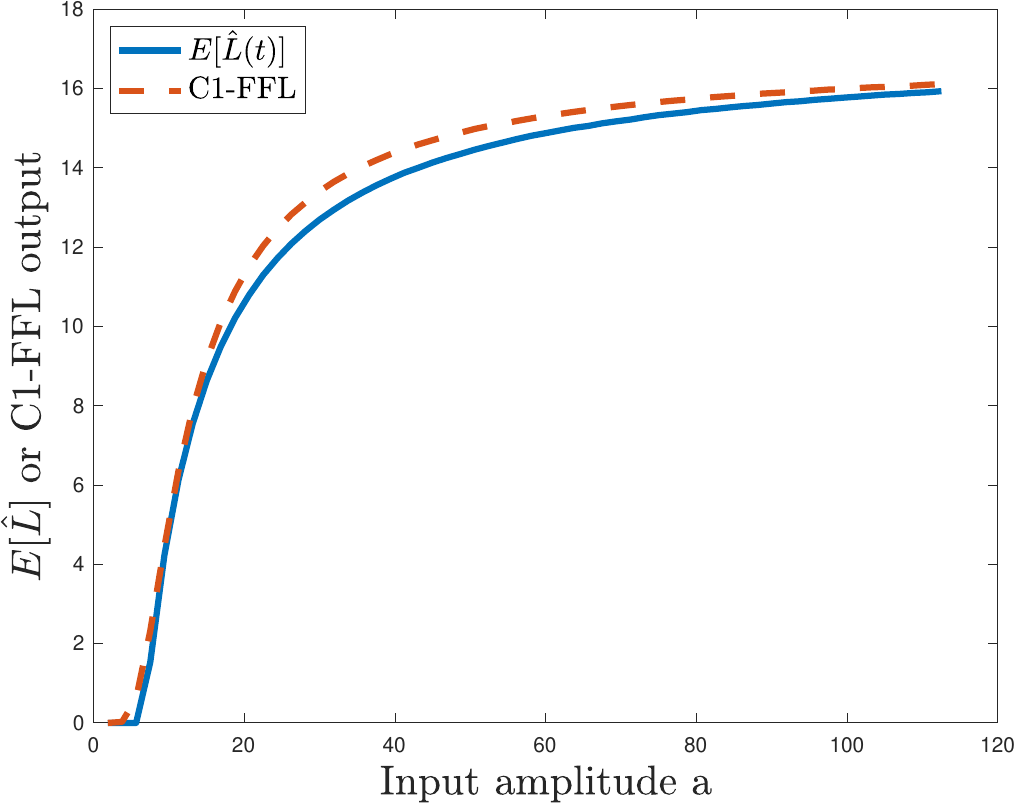}
        \caption{}
        \label{fig:opt_a}
    \end{subfigure}   
    
         \begin{subfigure}[t]{0.32\textwidth}
        \centering
        \includegraphics[scale=0.28]{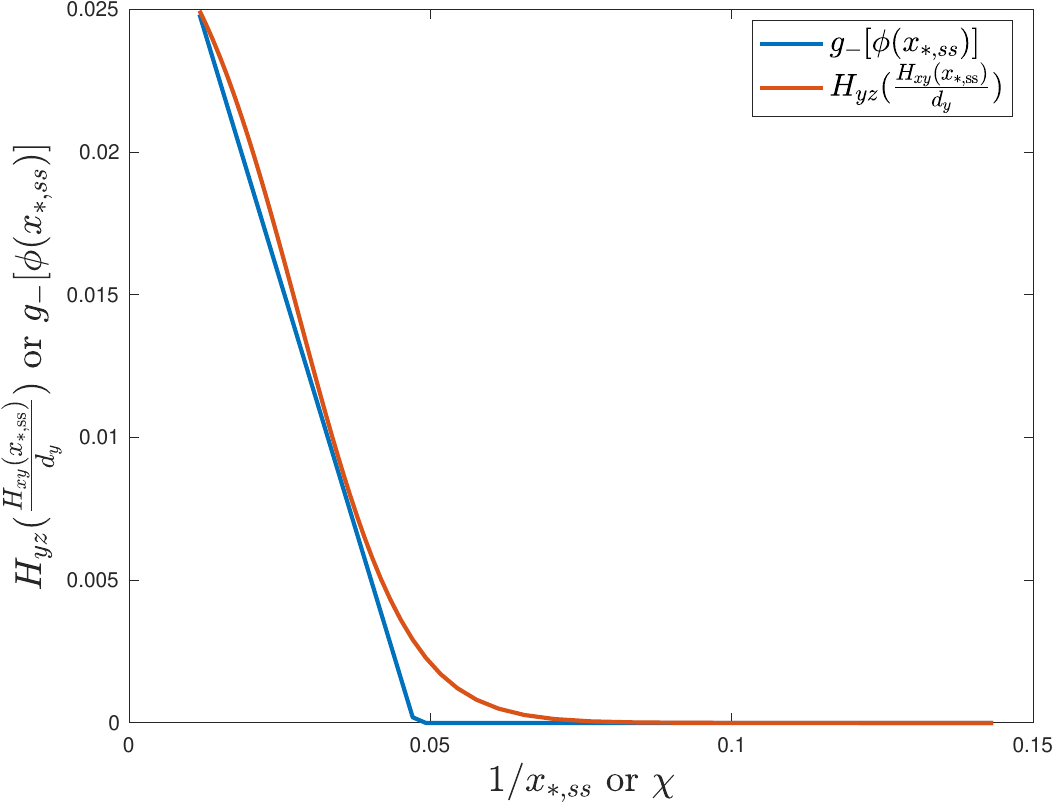}
        \caption{}
        \label{fig:match:opt}
    \end{subfigure}    
     \begin{subfigure}[t]{0.32\textwidth}
        \centering
        \includegraphics[scale=0.28]{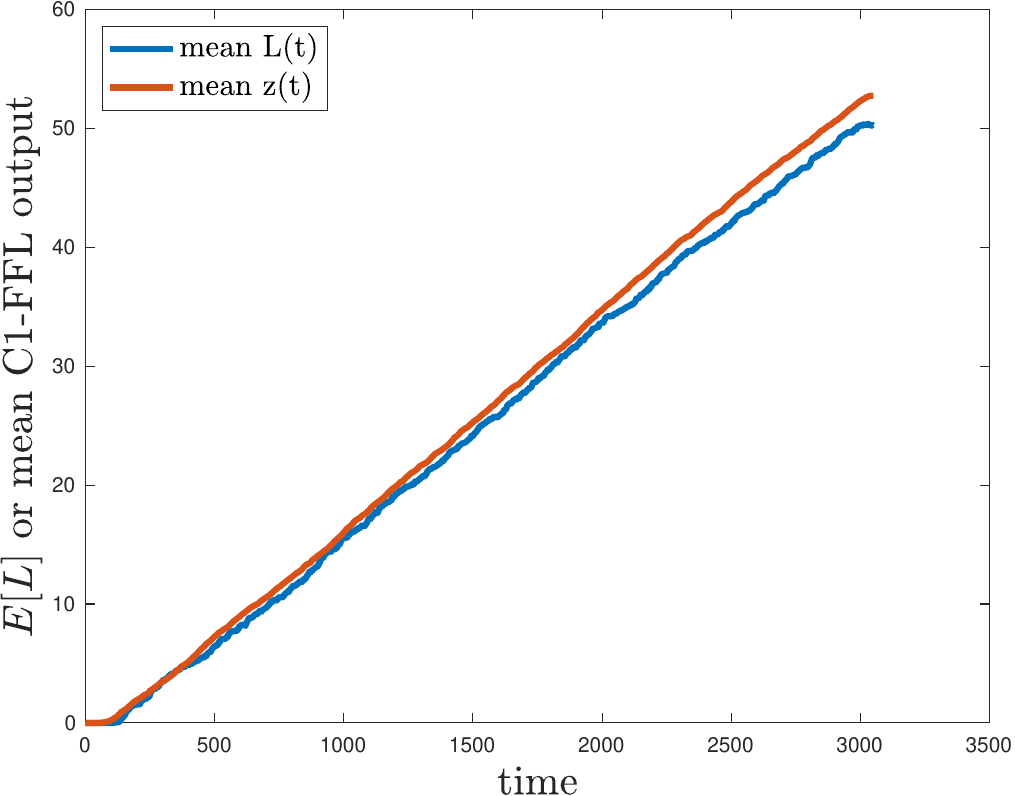}
        \caption{}
        \label{fig:opt_tau}
    \end{subfigure}       
\caption{Numerical results for Sec.~\ref{sec:c1ffl}. (a) Comparing the C1-FFL output against ${\rm E}[\hat{L}(t)]$ for three different values of $\alpha$. (b) Compare $z(t)$ (C1-FFL) and $\hat{L}(t)$ at time $t = 800$ for $a \in [12.5,100]$. 
(c) Comparing $g_- \;  [\phi(x_{*,{\rm ss}})]_+$ and $H_{yz}(\frac{H_{xy}(x_{*,{\rm ss}})}{d_y})$. Note the horizontal axis is $\chi = \frac{1}{x_{*,{\rm ss}}}$
(d) Comparing the mean C1-FFL output from tau-leaping against mean $L(t)$ from SSA. 
}
\label{fig:c1ffl_fit}
\end{figure*}

This numerical example uses the first set of parameter values as in Sec.~\ref{sec:int_approx:numer}. Our aim is to show that we can choose a set of C1-FFL parameters so that $z(t) \approx \hat{L}(t)$ for a range of input amplitude $\alpha$. We do this by fixing $H_{xz}()$ as $\frac{N g_+ x_*(t)}{g_+ x_*(t) + g_-}$, and by fitting the C1-FFL parameters in $H_{xy}()$ and $H_{yz}()$ in \eqref{eq:ffl_all} by nonlinear optimisation. The data for fitting is obtained from varying $\alpha$ from 3.125 to 62.5. For each $\alpha$, we compute ${\rm E}[\hat{L}(t)]$ using \eqref{eq:Lfinal_mean} and use them to fit the parameters of the C1-FFL. The fitted values of the C1-FFL parameters are: $k_{xy} = 9.66$, $n_{xy} = 1.65$, $K_{xy} = 4.36$, $d_y = 0.037$, $h_{yz} = 0.135$,  $n_{yz} = 50.0$ and $K_{yz} = 255.4$.   

Fig.~\ref{fig:opt_c1ffl} compares ${\rm E}[L(t)]$ and the C1-FFL output $z(t)$ for three different values of $\alpha$: 7.5, 30 and 75. It can be seen that they match very well.  Next, we compare the value of $z(t)$ and $\hat{L}(t)$ at time $t = 800$ for $\alpha \in [1.9,112]$. Fig.~\ref{fig:opt_a} shows that the match is good for a large range of $\alpha$.   

Although we use optimisation to obtain the parameters of the Hill functions $H_{xy}()$ and $H_{yz}()$. We find that their values are compatible to our intuitive argument in Sec.~\ref{sec:hatL:c1ffl}. First, we say that $K_{yz}$ should be chosen close to the value of $y(d_0)$ for a range of input amplitudes. For input amplitudes of 5, 15 and 480 (which correspond to a mean probability of 0.4, 0.6 and 0.8 that the promoter $\widetilde{\rm Z}_{X_*}$ is bound), the values of $y(d_0)$ are respectively, 229.2, 247.3 and 252.9. These values are pretty close to $K_{yz} = 255.4$ obtained from using optimization. Second, the steady state value of $y(t)$ is around 259 for a range of input amplitudes and $y(d_0)$ has not yet reached the steady state. Third, we argue that $H_{yz}(\frac{H_{xy}(x_{*,{\rm ss}})}{d_y})$ and $g_- \;  [\phi(x_{*,{\rm ss}})]_+$ should be approximately equal, i.e. \eqref{eq:match:xss} or \eqref{eq:match:chi}. Fig.~\ref{fig:match:opt} plots the two sides of \eqref{eq:match:chi} and shows that they match well except near the turning point. Fourth, the optimised value of $n_{xy}$ is 1.65 which is similar to the predicted value of 1. 

We next study the behaviour of the fitted C1-FFL when $s(t)$ in \eqref{eq:ffl_all} is the stochastic persistent input used in Sec.~\ref{sec:int_approx:numer}. We apply the tau-leaping simulation algorithm in \cite{Tian:2006dz} to simulate the stochastic behaviour of the C1-FFL from its ODE model. We perform 100 rounds of simulation to obtain the mean of the C1-FFL output $z(t)$. We compare the mean of $z(t)$ against the mean log-likelihood computed in Sec.~\ref{sec:int_approx:numer} in Fig.~\ref{fig:opt_tau} and we can see that they match fairly well. Next, we use $z(t)$ to study the TP (true positive) and FP (false positive) rates in the same way as in Sec.~\ref{sec:int_approx:tpfp}. Both persistent and transient inputs are used in this study, and 100 simulation runs are performed for each type of input. The magenta line in Fig.~\ref{fig:tp} (Page~\pageref{fig:tp}) shows how the TP rates vary with the decision threshold. It can be seen that the TP curve for the C1-FFL is in between those for $L(t)$ and $\hat{L}(t)$. We find that we can again find a range of positive thresholds such that the TP rate is high and FP rate is zero. 

\begin{remark}
\label{re:nonzero_dz} 
We have shown that on the condition that $d_z$ in \eqref{eq:ffl3} is zero, then we can find parameters of the C1-FFL such that $z(t)$ in \eqref{eq:ffl_all} is  approximately equal to ${\rm E}[\hat{L}(t)]$ in \eqref{eq:Lfinal_mean}. Let us now assume $d_z$ is non-zero and we add the term $-d_z {\rm E}[\hat{L}(t)]$ to the RHS of \eqref{eq:Lfinal_mean}, then we can use the same C1-FFL parameters as before to make $z(t)$ in \eqref{eq:ffl_all} to be approximately equal the new ${\rm E}[\hat{L}(t)]$. 
\end{remark} 

\begin{remark}
\label{re:v} The derivations in Secs.\ref{sec:results}-\ref{sec:c1ffl} assume that the reactions take place in a volume $V$ of 1. If the volume is not 1, \eqref{eq:Lfinal_mean} remains the same. Note that both $g_-$ and the term inside $[ \; ]_-$ in \eqref{eq:Lfinal_mean} are independent of $V$. If we divide both sides of \eqref{eq:Lfinal_mean} by $V$, we have $\frac{d({\rm E}[\hat{L}(t)]/V)}{dt} = \; {\rm E}[\widetilde{z}_{X_*}(t)] \; g_- \; \pi(t) \;  [...]_-$. Thus, the argument in Sec.~\ref{sec:hatL:c1ffl} is still valid if we interpret the output of the C1-FFL as log-likelihood ratio per volume. 
\end{remark}

\begin{remark}
Our interpretation of the C1-FFL as a statistical persistence detector shows an interesting signal processing architecture involving parallel processing. The C1-FFL has two arms. The short arm produces the signals ${X_*}(t)$ and $\widetilde{Z}_{X_*}(t)$ which contain information on whether the inducer signal is persistent or not. The longer arm then makes use of the signals ${X_*}(t)$ and $\widetilde{Z}_{X_*}(t)$ to approximately compute, in a parallel manner, the approximate log-likelihood ratio. It is also interesting to see that the gene $\widetilde{\rm Z}$, with its two promoter sites, has the dual roles of generating the signal to be processed as well as processing the signal. 
\end{remark} 

\subsection{Discussion}
\revisedtext{Biochemical circuits are known to be impacted by both intrinsic and extrinsic noise {\cite{Swain:2002kn}}. An open problem is to design a C1-FFL, which is based on the mass-reaction kinetics, for persistence detection using the probabilistic framework studied in this paper. In particular, we envisage using this design problem to study the impact intrinsic noise (which comes from reaction kinetics) and extrinsic noise (which comes from input signal properties such as how the signal is generated, amplitude, duration etc.) on the detection and false alarm rates. } 

% We envisage this design problem should have two goals. The first goal is to ensure that the mean output of the C1-FFL is approximately equal to the mean log-likelihood of the persistence detection problem. We learn from this paper that this goal should be feasible. The second goal is to minimise the variance of the difference between the C1-FFL output and the mean log-likelihood. If we can achieve the second goal, then we can get a C1-FFL which has good statistical property. 

\revisedtext{Another open problem is to study how our work can be used in conjunction with the sequential probability ratio test (SPRT) {\cite{Wald:1945tg}}. The SPRT uses the log-likelihood ratio $L(t)$, such as that defined in {\eqref{eq:L}} which uses data up till time $t$, with two thresholds $\theta_1$ and $\theta_0$ where $\theta_1 > \theta_0$. At any point in time, if $L(t) > \theta_1$ or $L(t) < \theta_0$ is true, then a decision is reached (deciding for, respectively, ${\cal H}_1$ or ${\cal H}_1$); otherwise a decision has not been reached and data collection will continue. It is suggested in {\cite{Siggia:2013dd}} that the SPRT is useful for biological detection because theoretically SPRT has a shorter decision time. We can use our approximate log-like ratio $\hat{L}(t)$ in the spirit of SPRT. In our case, only one threshold $\theta_1 (> 0)$ will be used. At any point in time, if $\hat{L}(t) > \theta_1$ then a decision is reached for ${\cal H}_1$; otherwise a decision has not been reached. (Note the similarity to the study in Sec.~{\ref{sec:int_approx:tpfp}}.) This set up will be useful for practical applications where an action is needed if ${\cal H}_1$ is true but otherwise no actions are required. We envisage that this set up will have a shorter decision time as in SPRT. It will be interesting to study how the design parameters mentioned in the last paragraph will influence the decision time and accuracy. This study is important for practical implementation. 
}

\section{Conclusions}
\label{sec:final} 
This paper shows that the C1-FFL gene circuits can be used to approximately compute the log-likelihood ratio for statistical detection of persistent signals. \revisedtext{In order to derive this result, we use new methods to show how log-likelihood ratio can be approximately computed and to show how this approximation can be mapped to the parameters of a C1-FFL. An open problem is to study the impact of intrinsic and extrinsic noise on the performance of the C1-FFL based statistical detector.
Another open problem is to consider using the approximate log-likelihood ratio computation together with sequential probability ratio test to realise fast persistent detection.} 
Finally, note that the methodology in this paper can be used to derive biochemical circuits that can be used to solve other statistical signal processing problems.

\ifarxiv 
	%\bibliographystyle{IEEEtran} 
	%\bibliography{nano,book} 
	% Generated by IEEEtran.bst, version: 1.12 (2007/01/11)

\else
	%\bibliographystyle{IEEEtran}
	%\bibliography{nano,book} 
	
\fi

% \clearpage % To clear the floats 

\appendices

\ifarxiv 
	\section{Proof of (\ref{eq:L})}
\else
    \section{Deriving the ODE for the exact log-likelihood ratio}
\fi    	
\label{app:sol:dp}
Recalling that $\widetilde{\cal Z}_{X_*}(t)$ is the history of $\widetilde{Z}_{X_*}(t)$ in the time interval $[0,t]$. In order to derive (\ref{eq:L}), we consider the history $\widetilde{\cal Z}_{X_*}(t+\Delta t)$ as a concatenation of $\widetilde{\cal Z}_{X_*}(t)$ and $\widetilde{Z}_{X_*}(t)$ in the time interval $(t,t+\Delta t]$. We assume that $\Delta t$ is chosen small enough so that no more than one reaction can take place in $(t, t+\Delta t]$. Given this assumption and right continuity of continuous-time Markov Chains, we can use $\widetilde{Z}_{X_*}(t+\Delta t)$ to denote the history of $\widetilde{Z}_{X_*}(t)$ in $(t,t+\Delta t]$. 

Consider the likelihood of observing the history $\widetilde{\cal Z}_{X_*}(t+\Delta t)$ given hypothesis ${\cal H}_i$: 
\begin{eqnarray}
&     & {\rm P}[\widetilde{\cal Z}_{X_*}(t+\Delta t) | {\cal H}_i]  \label{eqn:star:like:1} \\
& = & {\rm P}[\widetilde{\cal Z}_{X_*}(t) \; \mbox{\textsc{and}} \; \widetilde{Z}_{X_*}(t+\Delta t) | {\cal H}_i] \label{eqn:star:like:2} \\
& = & {\rm P}[\widetilde{\cal Z}_{X_*}(t) | {\cal H}_i]  \; {\rm P}[ \widetilde{Z}_{X_*}(t+\Delta t) | {\cal H}_i, \widetilde{\cal Z}_{X_*}(t)] \label{eqn:star:like:3} 
\end{eqnarray}
where we have expanded $\widetilde{\cal Z}_{X_*}(t+\Delta t)$ in \eqref{eqn:star:like:1} using concatenation.

By using \eqref{eqn:star:like:3} in the definition of log-likelihood ratio, we can show that: 
\begin{align} 
L(t+\Delta t) = L(t) + \log \left( \frac{{\rm P}[\widetilde{Z}_{X_*}(t+\Delta t) | {\cal H}_1, \widetilde{\cal Z}_{X_*}(t)]}{{\rm P}[\widetilde{Z}_{X_*}(t+\Delta t) | {\cal H}_0, \widetilde{\cal Z}_{X_*}(t)]} \right) \label{eq:app:L}
\end{align}

The condition probability ${\rm P}[\widetilde{Z}_{X_*}(t+\Delta t) | {\cal H}_i, \widetilde{\cal Z}_{X_*}(t)]$ is the prediction of the number of $\widetilde{\rm Z}_{X_*}$ molecules at time $t+\Delta t$ based on its history up till time $t$. This conditional probability can be obtained by solving a  Bayesian filtering problem over a continuous-time Markov chain which describes the dynamics of the chemical reactions in \eqref{cr:z_all} and those that produce \cee{S} \cite{Bronstein:2018eh}. 
We considered how this conditional probability could be evaluated in our earlier work \cite{Awan:2017fm}. The key result in \cite{Awan:2017fm} says that ${\rm P}[\widetilde{Z}_{X_*}(t+\Delta t) | {\cal H}_i, \widetilde{\cal Z}_{X_*}(t)]$ can be expressed in terms of the predicted rate of the chemical reactions that $\widetilde{\rm Z}_{X_*}$ are involved in. By using \cite{Awan:2017fm,Bronstein:2018eh}, we have: 
\begin{align}
&  {\rm P}[\widetilde{Z}_{X_*}(t+\Delta t) | {\cal H}_i, \widetilde{\cal Z}_{X_*}(t)] =\nonumber \\ 
& \delta_{\widetilde{Z}_{X_*}(t+\Delta t),\widetilde{Z}_{X_*}(t) + 1} \; g_+ (N - \widetilde{Z}_{X_*}(t))  \;  J_i(t_-) \; \Delta t +  \nonumber \\ 
& \delta_{\widetilde{Z}_{X_*}(t+\Delta t),\widetilde{Z}_{X_*}(t) - 1} \; g_- \widetilde{Z}_{X_*}(t) \; \Delta t \; + \nonumber  \\
& \delta_{\widetilde{Z}_{X_*}(t+\Delta t),\widetilde{Z}_{X_*}(t)}  \times \nonumber \\ 
& (1 - g_+ (N - \widetilde{Z}_{X_*}(t)) J_i(t) \; \Delta t - g_- \widetilde{Z}_{X_*}(t)  \; \Delta t) \label{eq:app:predictb}
\end{align}
where $\delta_{a,b}$ is the Kronecker delta which is 1 when $a$ equals to $b$ and zero otherwise, and  $J_i(t) = {\rm E}[X_*(t) | {\cal H}_i, \widetilde{\cal Z}_{X_*}(t)]$ is the expected number of \cee{X_*} molecules at time $t$ given Hypothesis $i$ and the history $\widetilde{\cal Z}_{X_*}(t)$.

Note that ${\rm P}[\widetilde{Z}_{X_*}(t+\Delta t) | {\cal H}_i, \widetilde{\cal Z}_{X_*}(t)]$ in \eqref{eq:app:predictb} is a sum of three terms with multipliers $\delta_{\widetilde{Z}_{X_*}(t+\Delta t),\widetilde{Z}_{X_*}(t) + 1}$, $\delta_{\widetilde{Z}_{X_*}(t+\Delta t),\widetilde{Z}_{X_*}(t) - 1}$ and $\delta_{\widetilde{Z}_{X_*}(t+\Delta t),\widetilde{Z}_{X_*}(t)}$. Since these multipliers are mutually exclusive, we have:
\begin{align}
 & \log \left( \frac{{\rm P}[\widetilde{Z}_{X_*}(t+\Delta t) | {\cal H}_1, \widetilde{Z}_{X_*}(t)]}{{\rm P}[\widetilde{Z}_{X_*}(t+\Delta t) | {\cal H}_0, \widetilde{Z}_{X_*}(t)]} \right)   \nonumber  \\
 = & 
 \delta_{\widetilde{Z}_{X_*}(t+\Delta t),\widetilde{Z}_{X_*}(t) + 1} \log \left( \frac{g_+ (N - \widetilde{Z}_{X_*}(t))  \;  J_1(t_-) \; \Delta t  }{g_+ (N - \widetilde{Z}_{X_*}(t))   \;  J_0(t_-) \; \Delta t }  \right) + \nonumber \\
 & \delta_{\widetilde{Z}_{X_*}(t+\Delta t),\widetilde{Z}_{X_*}(t) - 1} \log \left( \frac{g_- \widetilde{Z}_{X_*}(t) \; \Delta t}{g_- \widetilde{Z}_{X_*}(t) \; \Delta t}   \right) \nonumber + \\
 & \delta_{\widetilde{Z}_{X_*}(t+\Delta t),\widetilde{Z}_{X_*}(t)} \times \nonumber \\
& \log \left( \frac{ 1 - g_+ (N - \widetilde{Z}_{X_*}(t)) J_1(t) \; \Delta t - g_- \widetilde{Z}_{X_*}(t)  \; \Delta t}{ 1 - g_+ (N - \widetilde{Z}_{X_*}(t)) J_0(t) \; \Delta t - g_- \widetilde{Z}_{X_*}(t)  \; \Delta t }   \right) \nonumber \\
 \approx & 
\delta_{\widetilde{Z}_{X_*}(t+\Delta t),\widetilde{Z}_{X_*}(t) + 1} \log \left( \frac{J_1(t_-) }{J_0(t_-) }  \right) - \nonumber \\
& \delta_{\widetilde{Z}_{X_*}(t+\Delta t),\widetilde{Z}_{X_*}(t)} g_+ (N - \widetilde{Z}_{X_*}(t))  
\left( J_1(t) - J_0(t)  \right) \; \Delta t \label{eqn:app:likelihood} 
\end{align}
where we have used the approximation $\log(1 + f \; \Delta t) \approx f \; \Delta t$ and have ignored terms of order $(\Delta t)^2$ or higher to obtain \eqref{eqn:app:likelihood}. Note also that the above derivation assumes that $\frac{J_1(t)}{J_0(t)}$ is strictly positive so its logarithm is well defined; this can be achieved by proper choice of the hypotheses. 

By substituting \eqref{eqn:app:likelihood} into (\ref{eq:app:L}), we have after some manipulations and after taking the limit $\Delta t \rightarrow 0$:
\begin{align}
\frac{dL(t)}{dt} 
= & \lim_{\Delta t \rightarrow 0} \frac{\delta_{\widetilde{Z}_{X_*}(t+\Delta t),\widetilde{Z}_{X_*}(t) + 1} }{\Delta t}
\log \left( \frac{J_1(t_-)}{J_0(t_-)}  \right) - \nonumber \\
& \delta_{\widetilde{Z}_{X_*}(t+\Delta t),\widetilde{Z}_{X_*}(t)} g_+ (N - \widetilde{Z}_{X_*}(t))  
\left( J_1(t) - J_0(t)  \right) \label{eqn:app:logmapm1}  
\end{align}
In order to obtain (\ref{eq:L}), we use the following reasonings. First, the term $\lim_{\Delta t \rightarrow 0} \frac{\delta_{\widetilde{Z}_{X_*}(t+\Delta t),\widetilde{Z}_{X_*}(t) + 1} }{\Delta t}$ is a Dirac delta at the time instant that an \cee{X_*} molecule binds with $\widetilde{\rm Z}$ to form a $\widetilde{\rm Z}_{{\rm X}_*}$. Since the binding instants are also the times at which $\widetilde{Z}_{X_*}(t)$ jumps by $+1$, we can identify this term with $\left[ \frac{d\widetilde{Z}_{X_*}(t)}{dt} \right]_+$ where $[w]_+ = \max (w,0)$. Second, the term $\delta_{\widetilde{Z}_{X_*}(t+\Delta t),\widetilde{Z}_{X_*}(t)}$ is only zero when the number of $\widetilde{\rm Z}_{{\rm X}_*}$ molecule changes but the number of such changes is countable. In other words, $\delta_{\widetilde{Z}_{X_*}(t+\Delta t),\widetilde{Z}_{X_*}(t)}=1$ with probability one. This allows us to drop $\delta_{\widetilde{Z}_{X_*}(t+\Delta t),\widetilde{Z}_{X_*}(t)}$. Hence (\ref{eq:L}).  We remark that a more general framework of deriving log-likelihood ratio and log-posteriori probability in the reaction-diffusion master equation framework can be found in \cite{Chou:gc,Awan:2017fm,Chou:2015ga}.

\ifarxiv
	\section{Derivation of \eqref{eq:Lfinal} }
\else
    \section{Derivation of the intermediate approximation}
\fi    	
\label{app:ia}
The aim of this section is to show that the log-likelihood ratio computation in \eqref {eq:L} can be approximated by the intermediate approximation in \eqref{eq:Lfinal} for persistent signals in the time interval $[0,\min(d,d_1)]$. The derivation is based on the assumptions stated in Sec.~\ref{sec:assumptions}. We have divided the derivation into three steps.  \\

\begin{figure}[t]
    \centering
    \includegraphics[scale=0.5]{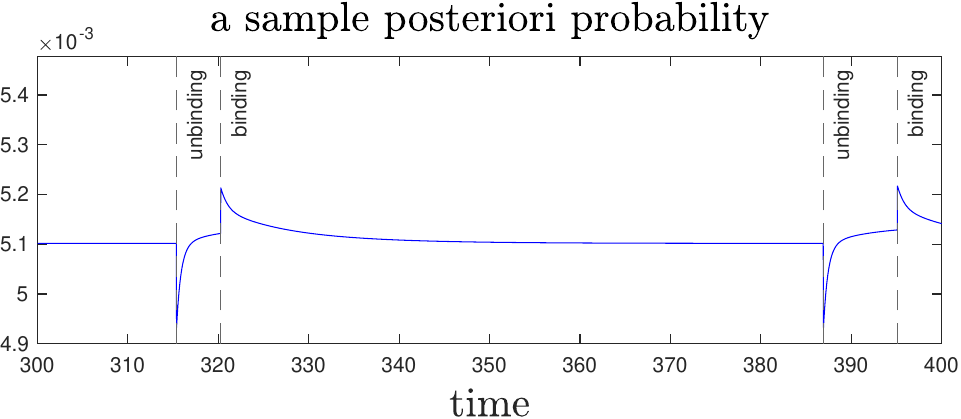}    
\caption{Typical temporal behaviour of a $\hat{p}_t(s,x_*)$.}
\label{fig:bayes:jumps}
\end{figure}

\begin{figure*}[t]
    \centering
    \begin{subfigure}[t]{0.32\textwidth}
        \centering
        \includegraphics[scale=0.29]{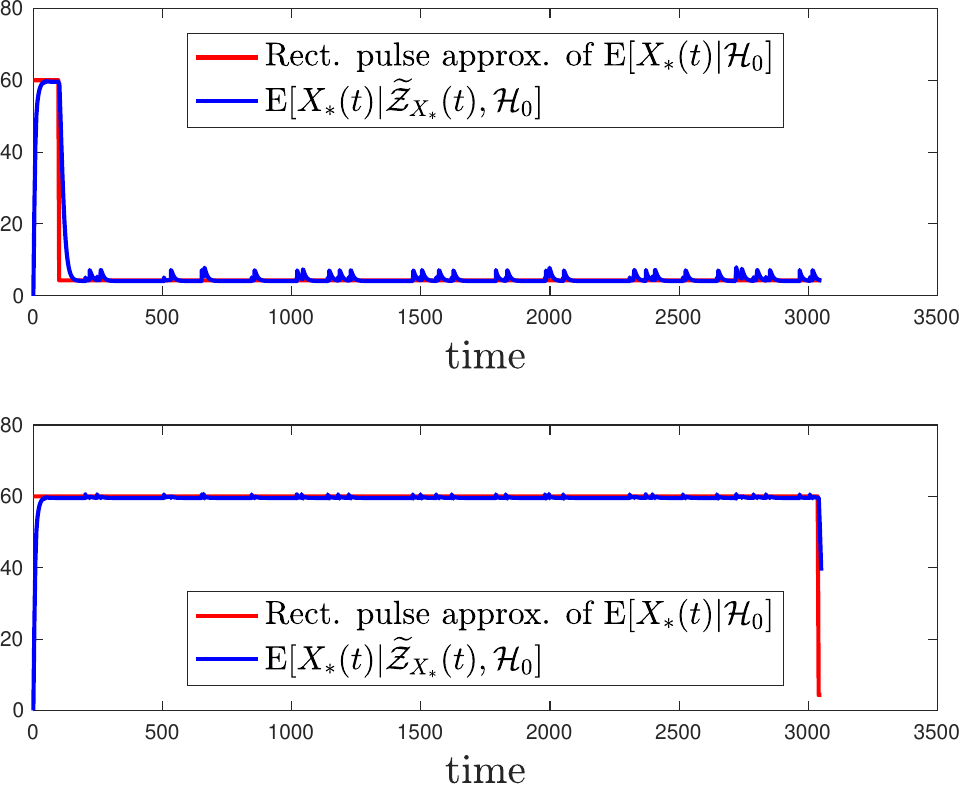}
        \caption{}
      \label{fig:app:filtering_A}
    \end{subfigure} 
    \begin{subfigure}[t]{0.32\textwidth}
        \centering
        \includegraphics[scale=0.29]{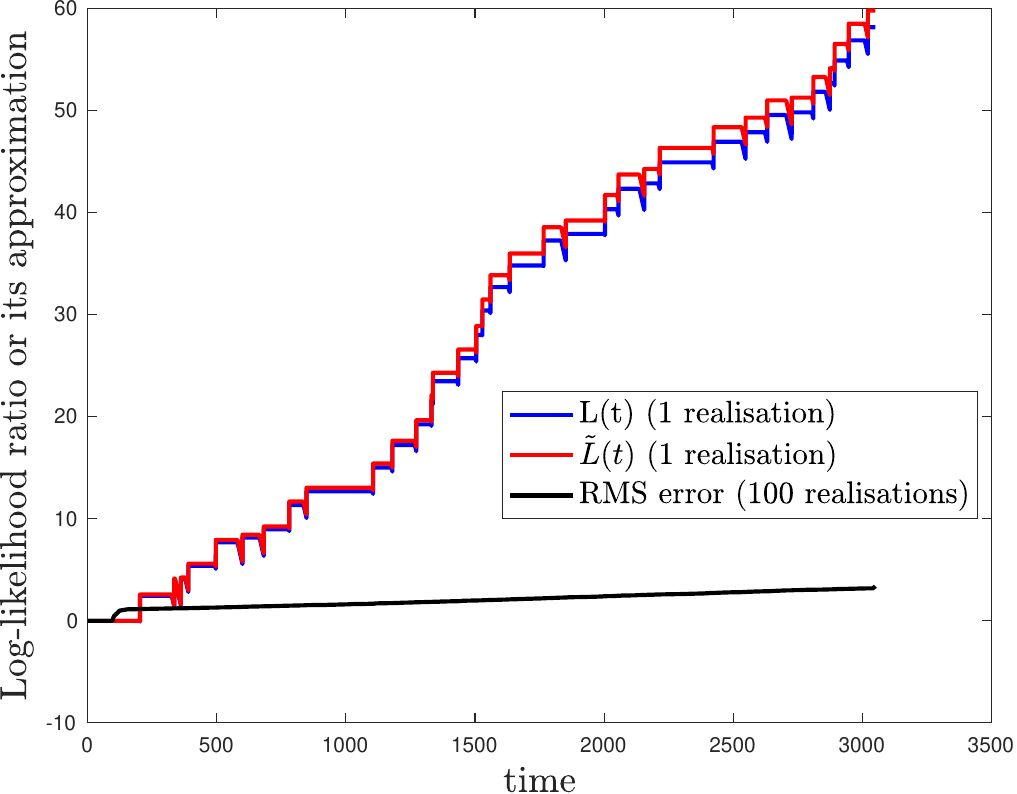}
        \caption{}
        \label{fig:app:filtering_B}
    \end{subfigure}   
    \begin{subfigure}[t]{0.32\textwidth}
        \centering
        \includegraphics[scale=0.29]{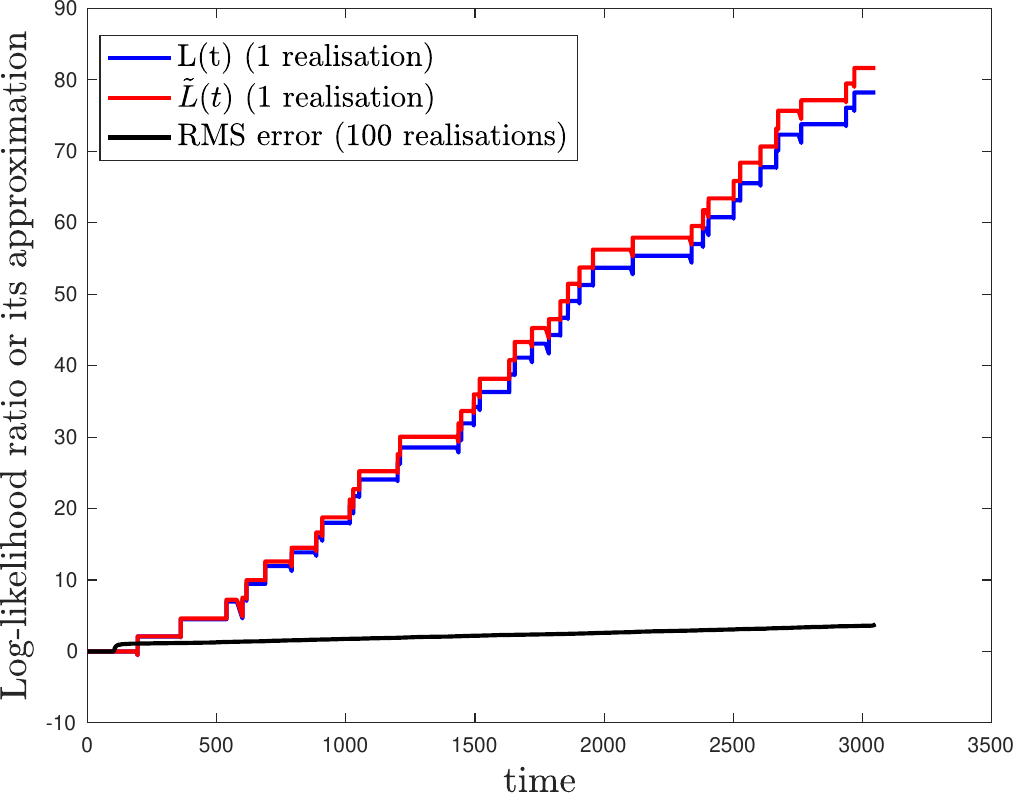}
        \caption{}
        \label{fig:app:filtering_C}
    \end{subfigure}    
\caption{(a) Demonstrating the accuracy of \eqref{eq:app:mean_xstar_0}. Top plot: ${\cal H}_0$; Bottom plot: ${\cal H}_1$. (b) and (c) Comparing  exact  log-likelihood ratio $L(t)$ \eqref{eq:L} and the approximation $\tilde{L}(t)$ \eqref{eq:app:L:2}. For (b),  $S_{\rm ref}^{\rm ON,ref} = 101$, and for (c), $S_{\rm ref}^{\rm ON,ref} = 270$.}
\label{fig:app:filtering}
\end{figure*}

\noindent{\bf (Step 1)} The aim of this step is to approximate ${\rm E}[X_*(t) | {\widetilde{\cal Z}}_{X_*}(t), {\cal H}_i]$ in \eqref {eq:L}. The computation of this expectation requires the solution of a Bayesian filtering problem which is computationally intensive. We will argue that if $g_+$ and $g_-$ are small, then we can use the approximation:
\begin{eqnarray}
{\rm E}[X_*(t) | {\widetilde{\cal Z}}_{X_*}(t), {\cal H}_i] &  \approx & 
\left\{
\begin{array}{cl}
X_1 & \mbox{ for } 0 \leq t < d_i  \\
X_0 & \mbox{otherwise}   \\
\end{array}
\right. 
. 
\label{eq:app:filtering} 
\label{eq:app:mean_xstar_0} 
\end{eqnarray}
where $i = 0,1$, and $X_0$ and $X_1$ are defined in \eqref{eq:ia:Xi}. In other words, we approximate ${\rm E}[X_*(t) |  {\widetilde{\cal Z}}_{X_*}(t), {\cal H}_i]$ by a rectangular pulse. 

In order to argue for \eqref{eq:app:mean_xstar_0}, we start by deriving the solution of the Bayesian filtering problem of using the history ${\widetilde{\cal Z}}_{X_*}(t)$ to determine the posteriori probability distribution $\hat{p}_t(s,x_*) \triangleq {\rm P}[S(t) = x, X_*(t) = x_* | {\widetilde{\cal Z}}_{X_*}(t), {\cal H}_i]$. The typical temporal behaviour of $\hat{p}_t(s,x_*)$ is depicted in Fig.~\ref{fig:bayes:jumps}. We can see that at the time instants where $\widetilde{\rm Z}$ binds or $\widetilde{\rm Z}_{X_*}$ unbinds, the probability $\hat{p}_t(s,x_*)$ has a discrete jump; at other times, i.e. between two consecutive jumps, the probability $\hat{p}_t(s,x_*)$ varies continuously. By using \cite[Eq. (21)]{Bronstein:2018eh}, we can show that the time evolution of the continuously varying part of $\hat{p}_t(s,x_*)$ is governed by the following ODE: 

\begin{align}
& \frac{d\hat{p}_t(s,x_*)}{dt} \nonumber \\
=&  [{\cal L}_i\hat{p}_t](s,x_*) \nonumber \\ 
&  - g_+ (x_* - {\rm E}[X_*(t) | {\widetilde{\cal Z}}_{X_*}(t), {\cal H}_i]) \; (N -  {\widetilde{\cal Z}}_{X_*}(t)) 
\label{eq:app:ia:full_filtering}
\end{align}
where
\begin{align}
& [{\cal L}_i\hat{p}_t](s,x_*) \nonumber \\
=& f_+ \; R_i(t) \; \hat{p}(s-1,x_*) - f_+ \; R_i(t) \; \hat{p}_t(s,x_*) + \nonumber \\ 
&  f_- \; (s+1) \; \hat{p}(s+1,x_*) - f_- \; s \; \hat{p}_t(s,x_*) +\nonumber \\ 
& k_+ \; s \; (M - x_* + 1) \; \hat{p}(s,x_*-1) -  \nonumber \\ 
& k_+ \; s \; (M - x_*)  \; \hat{p}_t(s,x_*) + \nonumber \\ 
& k_- \; (x_*+1) \; \hat{p}(s,x_*+1) - k_- \; x_* \; \hat{p}_t(s,x_*) 
\end{align}

If $g_+$ is small, then the last term in \eqref{eq:app:ia:full_filtering} can be viewed as a perturbation to the ODE:
\begin{align}
\frac{dp_t(s,x_*)}{dt} =  [{\cal L}_i p_t](s,x_*) 
\label{eq:app:ia:open}
\end{align}
This ODE is in fact the chemical master equation which governs the reactions \eqref{cr:S_prod}, \eqref{cr:z_all} with $g_+ = g_- = 0$. For small $g_+$, we have $\hat{p}_t(s,x_*) \approx p_t(s,x_*)$. In addition, if both $g_+$ and $g_-$ are small, then the time between two jumps in  $\hat{p}_t(s,x_*)$ is long; this further implies that $\hat{p}_t(s,x_*)$ is mostly around steady state, e.g. see Fig.~\ref{fig:bayes:jumps} in the time interval [320,386]. Overall, this means that we can approximate ${\rm E}[X_*(t) | {\widetilde{\cal Z}}_{X_*}(t), {\cal H}_i]$ by using the steady state mean number of \cee{X_*} molecules assuming $g_+ = g_- = 0$. When $g_+ = g_- = 0$, the steady state mean number of \cee{X_*} molecules is approximately given by the expression in \eqref{eq:ia:Xi}. Hence \eqref{eq:app:filtering}.  

After using the approximation in Step 1, \eqref {eq:L} becomes:
\begin{align}
 \frac{d\tilde{L}(t)}{dt} \approx& \left[ \frac{d\widetilde{Z}_{X_*}(t)}{dt} \right]_+ \log\left(\frac{X_1}{X_0} \right)  \; \pi(t) -  \nonumber  \\ 
 & g_+ (N - \widetilde{Z}_{X_*}(t)) (X_1-X_0)  \; \pi(t)  \label{eq:app:L:2} 
\end{align}
where $\pi(t)$ is defined in \eqref{eq:pi}. 

We now demonstrate the accuracy of \eqref{eq:app:filtering} and \eqref{eq:app:L:2} using numerical examples. We use the same parameter values as in Sec.~\ref{sec:int_approx:numer}. Fig.~\ref{fig:app:filtering_A} plots the two sides of \eqref{eq:app:filtering} for the two hypotheses. It can seen that, other than the transients, the approximation is fairly accurate. Next, we use Fig.~\ref{fig:app:filtering_B} to demonstrate the accuracy of \eqref{eq:app:L:2}. By using one realisation of $\widetilde{Z}_{X_*}(t)$, we calculate the exact  log-likelihood ratio $L(t)$ \eqref{eq:L} and the approximation $\tilde{L}(t)$ \eqref{eq:app:L:2}. The figure shows that the approximation is accurate. In addition, the figure also shows the RMS error between $L(t) - \tilde{L}(t)$ for 100 realisations of $\widetilde{Z}_{X_*}(t)$. It can be seem that the RMS error is small. In order to show that the approximation holds for different parameter settings, we use a different value  $S_{\rm ref}^{\rm ON,ref}$ and plot the result in  Fig.~\ref{fig:app:filtering_C}. 

\noindent{\bf (Step 2)} 
The aim of this step is to replace $\left[ \frac{d\widetilde{Z}_{X_*}(t)}{dt} \right]_+$ and $\widetilde{Z}_{X_*}(t)$ in \eqref{eq:app:L:2} by alternative expressions. Since $\pi(t)$ in \eqref{eq:app:L:2} is zero for $t < d_0$, we only have to consider input signals whose duration $d \geq d_0$. Recall that we assume in Sec.~\ref{sec:assumptions} that the pathway \eqref{cr:z_all} reaches steady state by $d_0$ if the input has a duration of at least $d_0$. This means that the probability distributions of \cee{X_*} and $\widetilde{Z}_{X_*}$ are in steady state in the time interval $[d_0,d]$

Note that the RHS of \eqref{eq:app:L:2} is the sum of two terms that do not depend on $L(t)$. We can therefore consider the contribution of each term to $L(t)$ separately. 

First, we consider the contribution of the first term on the RHS of \eqref{eq:app:L:2} to $L(t)$, which we will call $L_1(t)$: 
\begin{align}
 \frac{dL_1(t)}{dt}  =& \left[ \frac{d\widetilde{Z}_{X_*}(t)}{dt} \right]_+ \log\left(\frac{X_1}{X_0} \right)  \; \pi(t)  \label{app:L1:1} 
\end{align}
By using a method similar to that in \cite{Chou:2019gf}, which is based on the renewal theorem \cite{Grimmett}, we can show that \eqref{app:L1:1} can be approximated by: 
%By integrating the above equation, we have $L_1(t) = 0$ for $t \in [0,d_0)$; and for 
%$t \in [d_0,\min(d,d_1)]$:
%\begin{align}
%L_1(t) =&  
%\log\left(\frac{X_1}{X_0} \right) \int_{d_0}^t \left[ \frac{d\widetilde{Z}_{X_*}(\tau)}{d\tau} \right]_+  \; d\tau  \label{app:L1:2} 
%\end{align}
%By the assumption that the reaction pathway \eqref{cr:z_all} reaches steady state by $d_0$, the above equation can be written as:
%\begin{align}
%L_1(t) \approx&  \log\left(\frac{X_1}{X_0} \right) R_{\widetilde{Z}_{X_*}} \; (t - d_0)  \label{app:L1:3} 
%\end{align}
%where $R_{\widetilde{Z}_{X_*}}$ is the mean rate at which the binding of \cee{X_*} to $\widetilde{\rm Z}$ (i.e.~reaction \eqref{cr:on2}) occurs. It can be shown that, at steady state, the mean binding rate $R_{\widetilde{Z}_{X_*}}$ is equal to the mean unbinding rate of ${\widetilde{\rm Z}_{X_*}}$, which is $g_-$ times the mean number of ${\widetilde{\rm Z}_{X_*}}$. By using ergodicity, we have for $t \in [d_0,\min(d,d_1)]$:
%\begin{align}
%L_1(t) \approx&  \log\left(\frac{X_1}{X_0} \right) \int_{d_0}^t  g_- \; \widetilde{Z}_{X_*}(\tau) \; d\tau  \label{app:L1:4} 
%\end{align}
%Therefore, in differential form, we have
\begin{align}
\frac{dL_1(t)}{dt}  \approx&  \; g_- \; \widetilde{Z}_{X_*}(t)  \log\left(\frac{X_1}{X_0} \right)  \; \pi(t)  \label{app:L1:5} 
\end{align}

Next, we consider the contribution of the second term on the RHS of \eqref{eq:app:L:2} to $L(t)$,  which we will call $L_2(t)$: 
\begin{align}
\frac{dL_2(t)}{dt}  =& \;  g_+ (N - \widetilde{Z}_{X_*}(t)) (X_1-X_0)  \; \pi(t)  \label{app:L2:1} 
\end{align}
By integrating the above equation, we have $L_2(t) = 0$ for  $t \in [0,d_0)$; for $t \in [d_0,\min(d,d_1)]$, we have:
\begin{align}
L_2(t) =& \; (X_1-X_0) \; g_+ \int_{d_0}^{t} (N - \widetilde{Z}_{X_*}(\tau)) \; d\tau \label{app:L2:2} 
\end{align}
Since the reaction pathway is in steady state in the time interval $[d_0,\min(d,d_1)]$, we can replace the time average in \eqref{app:L2:2} by its ensemble average. In this part, we will overload the symbol $\widetilde{Z}_{X_*}$ to use it to refer to the random variable of the number of $\widetilde{\rm Z}_{X_*}$ molecules at steady state. This should not cause any confusion because the meaning should be clear from the context. In addition, we will overload the symbol $X_*$ in the same way. With this overloading, the mean number of \cee{X_*} and $\widetilde{\rm Z}_{X_*}$ molecules at steady state are denoted by ${\rm E}[X_*]$ and ${\rm E}[\widetilde{Z}_{X_*}]$ respectively. We can now rewrite \eqref{app:L2:2} as:
\begin{align}
L_2(t) \approx& \; (X_1-X_0) \; g_+  (N - {\rm E}[\widetilde{Z}_{X_*}]) \;   (t-d_0) \label{app:L2:3} 
\end{align}
In order to be able to connect to the C1-FFL, we will need to replace the expression $ (N - {\rm E}[\widetilde{Z}_{X_*}])$ by a different expression. The derivation of this replacement expression requires a few auxiliary results. \\

\noindent{\sl (Auxiliary Result 1)} By considering the global balance of the steady state of the reaction pathway \eqref{cr:z_all}, we have:
\begin{align}
g_+ {\rm E}[(N - \widetilde{Z}_{X_*})  X_*] =& \; g_- {\rm E}[\widetilde{Z}_{X_*}]   \label{app:L2:aux:1} 
\end{align} \\

\noindent{\sl (Auxiliary Result 2)} If the amplitude of the input is sufficiently high, then the number of \cee{X_*} molecules can be approximately modelled as a binomial distribution with $M$ trials and a success probability of $\frac{k_+ \alpha}{k_+ \alpha + k_-}$. 

Consider a binomial distribution $B(Q; m, f)$ with parameters $m$ (number of trials) and $f$ (success probability), then for sufficiently large $m$ and $f$, we have 
\begin{align}
\frac{1}{{\rm E}[Q]} \approx& \; {\rm E}[I(\frac{1}{Q})] \label{app:L2:aux:3}
\end{align}
where 
\begin{align}
I(\frac{1}{q}) =&
\left\{
\begin{array}{cl}
0               & \mbox{for } q = 0          \\
\frac{1}{q} & \mbox{for } q \geq 1
\end{array} 
\right.
\label{app:L2:aux:4}
\end{align}
This result essentially says that the mean of the reciprocal of a binomial random variable (with $\frac{1}{0}$ excluded) is approximately equal to the reciprocal of the mean of the binomial random variable. If $f = 1$ and $m \geq 1$, the binomial distribution has a single outcome with a non-zero probability so \eqref{app:L2:aux:3} is exact. Intuitively, if a probability has a single modal distribution with a narrow spread, then \eqref{app:L2:aux:3} holds approximately. For $f = 0.1$, the relative error of using \eqref{app:L2:aux:3} is 3.21\% for $m = 300$ and drops to 1.87\% for $m = 500$. 
% Similarly, for $f = 0.3$, the relative error is 0.79\% for $m = 300$ and drops to 0.47\% for $m = 500$. 
In general, the approximation is better for large $m$ and $f$. \\

\noindent{\sl (Auxiliary Result 3)} Since the inducer-TF reactions are faster than the TF-gene reactions and $M \gg N$, we can show that:
\begin{align}
{\rm E}[(N - \widetilde{Z}_{X_*})  X_*] \approx& \; {\rm E}[(N -\widetilde{Z}_{X_*})]    \; {\rm E}[ X_*]   \label{app:L2:aux:3_1} 
\end{align}
We will argue that the above approximation holds by using time average to compute ${\rm E}[(N - \widetilde{Z}_{X_*})  X_*]$. Let $t_0$, $t_1$, $\ldots$ be a sequence of time instants at which $\widetilde{\rm Z}_{X_*}$ changes its value. Since the continuous-time Markov chain associated with the chemical system is ergodic, we have:
\begin{align}
{\rm E}[(N - \widetilde{Z}_{X_*}) X_{*} ] =&  
\sum_{i = 0}^{\infty}  (N - \widetilde{Z}_{X_*}(t_i)) \int_{t_i}^{t_{i+1}} X_*(t) \; dt 
\label{app:L2:aux:3x} 
\end{align}
Since the TF-gene reactions are slow in comparison, $\widetilde{\rm Z}_{X_*}$ is a slow species while \cee{X_*} is a fast species. This means the time interval $[t_i,t_{i+1})$ (during which the count of $\widetilde{\rm Z}_{X_*}$ is a constant) is likely to be long compared to the time-scale of the fast species \cee{X_*}. This allows us to approximate the integral on the RHS of \eqref{app:L2:aux:3x} by ${\rm E}[X_*] (t_{i+1} - t_i)$. Hence \eqref{app:L2:aux:3_1}. Note that the above argument is identical to the one used in \cite{Cao:2005gj} to derive the slow-scale tau-leaping simulation algorithm. \\

\noindent{\sl (Auxiliary Result 4)} By using the same argument as in Auxiliary Result 3, we can show that:
\begin{align}
{\rm E}[(N - \widetilde{Z}_{X_*}) I(\frac{1}{X_{*}})] \approx& \; {\rm E}[(N -\widetilde{Z}_{X_*})]    \; {\rm E}[I(\frac{1}{X_{*}})]   \label{app:L2:aux:3_2} 
\end{align}
\\

We will now use the above auxiliary results and \eqref{app:L2:3} to derive the replacement expression. By using Auxiliary Results 1 and 3, we have:
\begin{align}
g_+ {\rm E}[(N -\widetilde{Z}_{X_*})] \approx& \; g_- {\rm E}[\widetilde{Z}_{X_*}]  \frac{1}{{\rm E}[ X_*]}   \label{app:L2:aux:4a}
\end{align}
We then apply Auxiliary Result 2 to the RHS of \eqref{app:L2:aux:4a} to obtain:
\begin{align}
g_+ {\rm E}[(N -\widetilde{Z}_{X_*})] \approx& \; g_- {\rm E}[\widetilde{Z}_{X_*}] {\rm E}[I(\frac{1}{X_{*}})]   \label{app:L2:aux:5}
\end{align}
By applying Auxiliary Result 4 to the RHS of \eqref{app:L2:aux:5}, we have:
\begin{align}
g_+ {\rm E}[(N -\widetilde{Z}_{X_*})] \approx& \; g_- {\rm E}[ \widetilde{Z}_{X_*} I(\frac{1}{X_{*}}) ] \label{app:L2:aux:6}
\end{align}

By substituting \eqref{app:L2:aux:6} into \eqref{app:L2:3}, we have:
\begin{align}
L_2(t) \approx& \; (X_1-X_0) \; \;  g_-  {\rm E}[ \widetilde{Z}_{X_*} I(\frac{1}{X_{*}}) ] (t-d_0) \label{app:L2:4} 
\end{align}
By turning the above equation into the differential form, we have:
\begin{align}
\frac{dL_2(t)}{dt}  \approx& \;  (X_1-X_0)  g_-  \widetilde{Z}_{X_*}(t) I(\frac{1}{X_{*}(t)})  \; \pi(t)  \label{app:L2:5} 
\end{align}
 
Next, by combining \eqref{app:L1:5} and \eqref{app:L2:5}, we have:
\begin{align}
\frac{dL(t)}{dt}  \approx&  \; g_- \; \widetilde{Z}_{X_*}(t) \; \pi(t) \left\{ \log\left(\frac{X_1}{X_0} \right)    -  (X_1-X_0)  I(\frac{1}{X_{*}(t)})  \right\}
\label{app:L:3} 
\end{align}
 
\noindent{\bf (Step 3)} 
Since a set of chemical reactions can be modelled by a set of ODEs, we want to turn the ODE in \eqref{app:L:3} into a form that can be implemented by a set of chemical reactions. However, \eqref{app:L:3} cannot be directly implemented by chemical reactions because log-likelihood ratio can take both positive and negative values but chemical concentration is always non-negative. Although \cite{Oishi:2011ig} has derived a chemical computation system that can have both positive and negative numbers, it requires double the number of species and reactions. As in our previous work \cite{Chou:2019gf,Chou:2018jh}, we choose to compute only the log-likelihood ratio when it is positive. We do that by applying $[ \; ]_+$ to the RHS of \eqref{app:L:3}; we have:
\begin{align}
\frac{dL(t)}{dt}  \approx&  \; g_- \; \widetilde{Z}_{X_*}(t) \; \pi(t) \; \times \nonumber \\ 
     & \; \left[ \log\left(\frac{X_1}{X_0} \right)    -  (X_1-X_0)  I(\frac{1}{X_{*}(t)})  \right]_+
\label{app:L:4_0} 
\end{align}

We now replace $I(\frac{1}{X_{*}(t)})$ in  \eqref{app:L:4_0} by $\frac{1}{X_{*}(t)}$ to obtain:
\begin{align}
\frac{dL(t)}{dt}  \approx&  \; g_- \; \widetilde{Z}_{X_*}(t) \; \pi(t) \; \times \nonumber \\ 
     & \; \left[ \log\left(\frac{X_1}{X_0} \right)   -  (X_1-X_0)  \frac{1}{X_{*}(t)} \right]_+
\label{app:L:4} 
\end{align}
The removal of $I(\;)$ will not make much difference because the probability of having $X_*(t)$ equals to 0 is small when the input signal is persistent. Note that  \eqref{app:L:4} is the same as  \eqref{eq:Lfinal}. This completes the derivation for \eqref{eq:Lfinal}. \\

In order to derive \eqref{eq:Lfinal_mean}, we start from \eqref{app:L:4} and take expectation on both sides. If the amplitude $\alpha$ is sufficiently high, then there is a high probability that $X_*(t)$ is large. This means we can take the expectation operator to the inside of the $[ \; ]_+$ operator. After that we apply Auxiliary Results 2, 3 and 4 to obtain \eqref{eq:Lfinal_mean}.   \\

\ifarxiv
\section{Utility maximisation}
\label{app:util}
In this section, we will show that the utility maximisation problem \eqref{eq:max_util} leads to a detection criterion based on the likelihood ratio. The proof here uses the same method as Appendix 3A in \cite{Kay_v2} for proving the Neyman-Pearson lemma. 

By using Lagrangian multiplier $\lambda \geq 0$, we rewrite \eqref{eq:max_util} as the maximisation of:
\begin{eqnarray}
&    &  P_{TP} P_1 U_1 - \lambda  (P_{FP} P_0 C_1 + P_{TP}P_1 C_1 - C_{\rm max} ) \\
& = & (U_1 - \lambda C_1) P_1 P_{TP} - \lambda C_1 P_0 P_{FP} + \lambda C_{\rm max} \label{eq:util:lagmax}
\end{eqnarray}

Let ${\cal O}$ be the observations that are available to the detection problem. Also let ${\cal A}_0$ and ${\cal A}_1$ be two disjoint sets which are to be used as the decision regions for the detection problem. In particular, the detection problem will decide for hypothesis ${\cal H}_1$ (resp. ${\cal H}_0$) if ${\cal O} \in {\cal A}_1$ (${\cal O} \in {\cal H}_0$). Given these definitions, we can write $P_{TP}$ and $P_{FP}$ as:
\begin{eqnarray}
P_{TP} =   \int_{{\cal O} \in {\cal A}_1}  {\rm P}[ {\cal O} | {\cal H}_1]  \; d {\cal O} \label{eq:util:p_tp} \\
P_{FP}  =  \int_{{\cal O} \in {\cal A}_1}  {\rm P}[ {\cal O} | {\cal H}_0]  \; d {\cal O} \label{eq:util:p_fp}
\end{eqnarray}
By substituting \eqref{eq:util:p_tp} and \eqref{eq:util:p_fp} into \eqref{eq:util:lagmax}, we have the objective to be maximised is:
\begin{eqnarray}
 \int_{{\cal O} \in {\cal A}_1} (U_1 - \lambda C_1) P_1 \;  {\rm P}[ {\cal O} | {\cal H}_1] - \lambda C_1 P_0 \; {\rm P}[ {\cal O} | {\cal H}_0]  \; d {\cal O} + \lambda C_{\rm max} \label{eq:util:lagmax2}
\end{eqnarray}
In order to maximise \eqref{eq:util:lagmax2}, we should include ${\cal O}$ in ${\cal A}_1$ if the integrand in \eqref{eq:util:lagmax2} is positive. In other words, ${\cal A}_1$ should be the set of all ${\cal O}$'s such that:
\begin{eqnarray}
(U_1 - \lambda C_1) P_1 \;  {\rm P}[ {\cal O} | {\cal H}_1] - \lambda C_1 P_0 \; {\rm P}[ {\cal O} | {\cal H}_0] > 0 \label{eq:util:lagmax3} 
\end{eqnarray}
If $U_1 - \lambda C_1 > 0$ then \eqref{eq:util:lagmax3} is equivalent to: 
\begin{eqnarray}
\frac{ {\rm P}[ {\cal O} | {\cal H}_1]}{ {\rm P}[ {\cal O} | {\cal H}_0]} > \frac{ \lambda C_1 P_0 }{ (U_1 - \lambda C_1) P_1 } 
\end{eqnarray}
This shows that we can use a criterion based on the likelihood ratio to maximise the utility. Note that if the requirement $U_1 - \lambda C_1 > 0$ holds, then a non-empty ${\cal A}_1$ may be found because the benefit gained (or utility) from deciding for ${\cal A}_1$ outweighs the cost required. Otherwise, if $U_1 - \lambda C_1 \leq 0$, \eqref{eq:util:lagmax3} suggests that ${\cal A}_1$ should be an empty set. 

Note that \eqref{eq:util:lagmax2} can be interpreted as the maximisation of $P_{TP}$ subject to an upper bound on $P_{FP}$, which is the same class of optimisation formulation that the Neyman-Pearson lemma considers. 

We remark that we can see from the above derivation that if both the mean utility and mean cost are linear in the probabilities $P_{TP}$, $P_{FP}$, $P_{TN}$ and $P_{FN}$, then the utility maximisation problem can be solved by using a criterion based on the likelihood ratio. We further remark that it is straightforward to generalise the above proof to the case with non-zero utility and cost. 
\fi

\ifarxiv
\section{Necessity of delay}
\label{app:delay}
We use the method of contradiction to argue that there must be a delay in the cloud if $d_0 > t_{\rm ss}$. Let us assume that there is a ``memoryless" function $\psi({\rm E}[X_*(t)])$ which can carry out the computation in the cloud, i.e. 
\begin{eqnarray}
\psi({\rm E}[X_*(t)]) = g_- \; \pi(t) \;  [\phi({\rm E}[X_*(t)])]_+ \label{eq:memoryless} 
\end{eqnarray} 
Since we assume that the pathway \eqref{cr:z_all} reaches the steady state by $d_0$, we can find time instants $t_1$ and $t_2$ where $t_{\rm ss}  \leq t_1 < d_0 < t_2 < d$ such that ${\rm E}[X_*(t)]$ is at steady state at both times $t_1$ and $t_2$, i.e. ${\rm E}[X_*(t_1)]  = {\rm E}[X_*(t_2)]$. Let us consider \eqref{eq:memoryless} at time instants $t_1$ and $t_2$, we have:
\begin{eqnarray}
\psi({\rm E}[X_*(t_1)]) & = & 0 \hspace{1cm} (\mbox{because } \pi(t_1) = 0) \\ 
\psi({\rm E}[X_*(t_2)]) & = & g_- \;  [\phi({\rm E}[X_{*,ss}])]_+ > 0 
\end{eqnarray} 
which is a contradiction because ${\rm E}[X_*(t_1)]  = {\rm E}[X_*(t_2)]$. This establishes that, if $d_0 > t_{\rm ss}$, there must be a delay in the cloud in Fig.~\ref{fig:compute:Lhat}. 
\fi

\ifarxiv
\else
\begin{IEEEbiography}[{\includegraphics[width=1in,height=1.25in,clip,keepaspectratio]{chou}}]%
{Chun Tung Chou}
is an Associate Professor at the School of Computer Science and Engineering, the University of New South Wales, Australia. He received the BA degree in Engineering Science from the University of Oxford, UK and the PhD degree in Control Engineering from the University of Cambridge, UK. He is/was on the editorial board of IEEE Transactions on Molecular, Biological, and Multi-Scale Communications; IEEE Wireless Communications Letters and Nano Communication Networks. His research interests are in molecular communication, molecular computing and pervasive computing.
\end{IEEEbiography}
\fi

\ifarxiv
\else
	\EOD
\fi

\end{document}